\documentclass[twocolumns]{aa} 

\usepackage{graphicx}
\usepackage{natbib}
\usepackage{amstext}
\bibpunct{(}{)}{;}{a}{}{,}

\usepackage{txfonts}
\usepackage{times,amssymb}

\usepackage{color}

\def\HII{\mbox{\ion{H}{II}}}
\def\FFeII{\mbox{[\ion{Fe}{II}]}}
\def\S{\mbox{{\fontfamily{phv}\selectfont S}}}

\def\mag{\ifmmode^{\rm m }\else$^{\rm m}$\fi}
\def\fracarcsec{\hbox{$.\!\!^{\rm s}$}}
\def\fracmag{\hbox{$.\!\!^{\rm m}$}}
\def\as{$\,^{\prime\prime}\,$}

\def\hh{\ifmmode^{\rm h}\else$^{\rm h}$\fi}
\def\mm{\ifmmode^{\rm m}\else$^{\rm m}$\fi}
\def\ss{\ifmmode^{\rm s}\else$^{\rm s}$\fi}
\def\deg{\ifmmode^\circ\else$^\circ $\fi}
\def\amin{\ifmmode^\prime\else$^\prime $\fi}
\def\rahm#1#2{#1\hh\ #2\mm\ }
\def\decdm#1#2{\ifmmode{#1}\else{$#1$}\fi\deg\ #2\amin\ }
\def\ras#1#2{#1\fracarcsec#2 }
\def\decs#1#2{#1\farcs#2 }

\def\dec#1#2#3{\ifmmode{#1}\else{$#1$}\fi\deg\ #2\amin\ #3\as\ }
\def\rab#1#2#3#4{#1\hh\ #2\mm\ #3\fracarcsec#4 }
\def\decb#1#2#3#4{\ifmmode{#1}\else{$#1$}\fi\deg\ #2\amin\ #3\farcs#4 }

\begin{document}

\title{Outflows from the high-mass protostars \object{NGC~7538} IRS1/2
  observed with bispectrum speckle interferometry\thanks{Observations reported
  here were obtained at the BTA 6m telescope of the Special Astrophysical
  Observatory and at the MMT Observatory, a joint facility of the Smithsonian
  Institution and the University of Arizona. In addition, this work is based
  in part on archival data obtained with the Spitzer Space Telescope, which is
  operated by the Jet Propulsion Laboratory, California Institute of
  Technology under a contract with NASA.}}
 \subtitle{Signatures of flow precession}

\titlerunning{Outflows from the high-mass protostars NGC~7538 IRS1/2}

\author{S.~Kraus\inst{1} \and Y.~Balega\inst{2} \and M.~Elitzur\inst{3}
  \and K.-H.~Hofmann\inst{1} \and Th.~Preibisch\inst{1}
  \and A. Rosen\inst{1} \and D.~Schertl\inst{1} \and G.~Weigelt\inst{1}
  \and E.~T.~Young\inst{4}
}

\authorrunning{Kraus et al.}

\offprints{skraus@mpifr-bonn.mpg.de}

\institute{Max Planck Institut f\"ur Radioastronomie, Auf dem H\"ugel 69,
  53121 Bonn, Germany
  \and
  Special Astrophysical Observatory, Russian Academy of Sciences, Nizhnij Arkhyz,
  Zelenchuk region, Karachai-Cherkesia, 357147, Russia
  \and
  Department of Physics \& Astronomy, University of Kentucky, Lexington,
  KY 40506, USA
  \and
  Steward Observatory, University of Arizona, 933 North Cherry Avenue,
  Tucson, AZ 85721, USA
}

\date{Received February 22, 2006; accepted April 10, 2006}

\abstract
{\object{NGC~7538~IRS1} is a high-mass ($30~M_{\sun}$) protostar with a CO
  outflow, an associated ultracompact \HII~region, and a linear methanol maser
  structure, which might trace a Keplerian-rotating circumstellar disk. The
  directions of the various associated axes are misaligned with each other.
}
{We investigate the near-infrared morphology of the source to clarify the
relations among the various axes.
}
{\textit{K'}-band bispectrum speckle interferometry was performed at two
6-meter-class telescopes---the BTA 6m telescope and the 6.5m MMT. Complementary IRAC
images from the \textit{Spitzer} Space Telescope Archive were used to relate
the structures detected with the outflow at larger scales.
}
{High-dynamic range images show 
fan-shaped outflow structure in which we detect 18 stars and several blobs of 
diffuse emission. We interpret the misalignment of various outflow axes in the
context of a disk precession model, including numerical hydrodynamic
simulations of the molecular emission. The precession period is $\sim
280$~years and its half-opening angle is $\sim 40^\circ$. A possible
triggering mechanism is non-coplanar tidal interaction of an (undiscovered)
close companion with the circumbinary protostellar disk. Our observations
resolve the nearby massive protostar \object{NGC~7538~IRS2} as a close binary
with separation of $195$~mas. We find indications for shock interaction
between the outflow activities in IRS1 and IRS2. Finally, we find prominent
sites of star formation at the interface between two bubble-like structures in
NGC~7538, suggestive of a triggered star formation scenario.
}
{ Indications of outflow precession have been discovered to date in a number of
massive protostars, all with large precession angles ($\sim$ 20--45$^\circ$).
This might explain the difference between the outflow widths in low- and
high-mass stars and add support to a common collimation mechanism.
}


\maketitle
%

\section{Introduction}

Protostellar disks and outflows are essential constituents of the star
formation process.  For high-mass protostellar objects (HMPOs), 
direct evidence for the presence of compact circumstellar disks is still rare,
whereas outflows seem to be omnipresent in the high-mass star forming regions.
Outflows remove not only angular momentum from the infalling matter, but also
help to overcome the radiation pressure limit to protostellar accretion, by
carving out optically thin cavities along which the radiation pressure can
escape~\citep{kru05}.

How outflows are collimated is a matter of ongoing debate and may depend on the
stellar mass of the outflow-driving source.  One of the arguments in support of
this conclusion is that outflows from high-mass stars appear less collimated
than the outflows and jets from their low-mass counterparts~\citep{wu04}.
Therefore, it has been suggested that outflows from HMPOs might be driven by strong
stellar winds, lacking a recollimation mechanism.  Since HMPOs typically form
in dense clusters, another possibility is confusion by the presence of
multiple collimated outflows.

However, since there is evidence that the binary frequency is significantly
higher for high-mass than for low-mass stars \citep[e.g., ][]{pre99}, another
possibility is that outflows from HMPOs simply appear wider, assuming they
undertake precession.  A few cases where outflow precession have been proposed
for HMPO outflows~\citep[e.g. ][]{she00, wei02, wei06} show precession angles
of $\sim 20$ to $45^\circ$; considerably wider than the jet precession angles
of typically just a few degrees observed towards low-mass
stars~\citep{ter99}. This is in agreement with the general picture that
high-mass stars form at high stellar density sites and therefore experience
strong tidal interaction from close companions and stellar encounters.

The detection of precessing jet-driven outflows from HMPOs adds support to the
hypothesis of a common formation mechanism for outflows from low to 
high-mass stars.  Furthermore, jet precession carries information about the
accretion properties of the driving source and, simultaneously, about
the kinematics and stellar population within its closest vicinity, yielding a
unique insight into the crowded places where high-mass star formation
occurs.\\

In this paper, we report another potential case of outflow precession
concerning the outflow from the high-mass~\citep[$30~M_{\sun}$, ][]{pes04}
protostellar object NGC~7538~IRS1.

We obtained bispectrum speckle interferometry of IRS1 and IRS2, which provides
us with the spatial resolution to study the inner parts of the outflow,
detecting filigreed fine structure within the flow.  Information about even
smaller scales is provided by the intriguing methanol maser feature, which was
detected at the position of this infrared source and which was modeled
successfully as a protostellar disk in Keplerian rotation~\citep{pes04}. To
search for outflow tracers on larger scales, we also present archival
\textit{Spitzer}/IRAC images.  In addition, this allows us to relate 
the sources studied with bispectrum speckle interferometry with the overall
star forming region and we find new hints for triggered star formation in this
region.\\

\subsection{Previous studies of NGC~7538}

The NGC~7538 molecular cloud is located in the Cas OB2 association in the
Perseus spiral arm at a distance of $\sim 2.8$~kpc~\citep{bli82}. Several
authors noted that NGC~7538 might present a case of triggered or induced star
formation since it shows ongoing star formation at various evolutionary
stages, apparently arranged in a northwest (most developed) to southeast
(youngest evolutionary stage) gradient~\citep{mcc91}. 

At optical wavelengths, the appearance of the region is dominated by diffuse
\HII~emission, which extends several arcminutes from the southeast to the
northwest~\citep{lyn86}. In 1974, \citeauthor*{wyn74} detected eleven infrared
sources (IRS1-11) in the NGC~7538 region, wherein IRS1--3 are located on the
southeast-corner of the fan-shaped \HII~emission in a small cluster of
OB-stars. IRS1 is the brightest NIR source within this cluster and is embedded
within an ultracompact (UC) \HII~region whose size was estimated to be $\sim
0\farcs4$~($n_e\approx10^5$~\mbox{cm$^{-3}$}, measured in 5 and 15~GHz CO
continuum, \citealt{cam84b}).  The spectral type was estimated to be
O7~\citep{aka05}, which implies a luminosity $\sim 9.6 \times 10^4
L_{\sun}$. VLA observations with a resolution down to $0\farcs1$ (=180~AU) also
revealed a double-peaked structure of ionized gas within the UC core (peaks
separated by $\sim 0\farcs2$), which was interpreted as a disk collimating a
north-south-oriented outflow~\citep{cam84b,gau95}.  This interpretation is also
supported by the detection of elongation of the dust-emitting region at
mid-infrared (MIR) wavelengths~(5~$\mu$m:~\citealt{hac82}; 11.7~$\mu$m and
18.3~$\mu$m:~\citealt{deb05}) and imaging studies performed in the
sub-millimeter continuum~\cite[350~$\mu$m, 450~$\mu$m, 800~$\mu$m, 850~$\mu$m,
1.3~mm:~][ showing an elliptical source with a size of $\sim 11\farcs6 \times
7\farcs6$ along PA\footnote{Following the convention, we  measure the
position angle (PA) from north to east.}~$\sim -80^\circ$]{san04} and CO line
emission~\cite[ showing a disk-like structure extending $\sim 22\arcsec$ in the
east-west direction]{sco86}. Also, polarization measurements of the infrared
emission around IRS1 can be construed in favor of the disk
interpretation~\citep{dyc78,tam91}. \citet{kaw92} carried out interferometric
CS (J=$2\rightarrow1$) observations and found a ring-like structure, which they
interpret as a nearly face-on protostellar disk of dense molecular gas.

Further evidence for outflow activity was found by~\citet{gau95}, who measured
the profile of the H66$\alpha$ recombination line and derived high velocities
of 250~\mbox{km~s$^{-1}$}, indicating a strong stellar outflow from IRS1. CO
(J=$1\rightarrow0$) spectral line mapping showed a bipolar flow~\citep{fis85}.
The mass outflow rate $\dot{M}_{\text{outflow}}$ from IRS1 was estimated to be
$\sim 5.4 \times 10^{-3}~M_{\sun}~\text{yr}^{-1}$~\citep{dav98}.
Interferometric observations by~\citet[ beam size 7\arcsec]{sco86} show that the
blue and red-shifted lobes are separated by 28\arcsec~with a position angle of
$-45^{\circ}$, and IRS1 is located on this axis just between the lobes of this
high-velocity ($-76$ to $-37$~\mbox{km~s$^{-1}$}) CO outflow. In comparing the
data obtained with various beam sizes~\citep{cam84b,kam89}, these seem to
indicate a change in the position angle of the flow direction at different
spatial scales, ranging from PA~$\sim 0^\circ$ at $0\farcs3$,  PA~$\sim
-25^\circ$ at 2\arcsec, PA~$\sim-35^\circ$ at 7\arcsec, to PA~$-40^\circ$
at 16\arcsec.

Within the immediate ($\sim 0\farcs5$) vicinity of IRS1, a large variety of masers
has been discovered, including OH~\citep{dic82},
H$_2$CO~\cite[formaldehyde,~][]{rot81, hof03},
NH$_3$~\cite[ammonia,~][]{mad86}, CH$_3$OH \cite[methanol, five features A, B,
C, D, E were detected at 6.7 and 12.2~GHz:~][]{men86, min98, min00},
$^{15}$NH$_3$~\citep{joh89}, and H$_2$O~\citep{kam90}.  Some of the masers show
only vague signs for a systematic alignment within linear ($^{15}$NH$_{3}$,
PA~$\sim-60^\circ$) or ring-like structures (H$_2$O, methanol-maser feature
E). However, the methanol-maser feature A represents one of the most convincing
cases of systematic alignment, in both linear spatial arrangement (PA~$\sim
-62^\circ$) and well-defined velocity gradient, observed to date in any maser
source. The qualitative interpretation of this structure as an edge-on 
circumstellar disk~\citep{min98} was later confirmed by the detailed modeling 
of \citet{pes04}, which showed that the alignment in the
position--line-of-sight (LOS) velocity diagram of maser feature A can be
modeled accurately assuming a protostellar disk with Keplerian rotation.

Aiming for a more complete picture, several authors
\citep[e.g.][]{min98,deb05} also tried to incorporate the presence of
methanol maser features B, C, D, and E in the circumstellar disk model for
feature A and interpreted them as part of an outflow which is oriented
perpedicular to feature A.  Since these maser features are southwards of the
putative circumstellar disk, it remains unclear why they appear blue-shifted
with respect to feature A~\citep{min98}, whereas the southern lobe of the
CO-outflow is red-shifted.

Besides the circumstellar disk interpretation for the origin of the maser
feature A mentioned above, an alternative scenario was proposed
by~\citet{deb05}, who suggested that feature A might trace the walls of an
outflow cavity.

The region was also intensively observed in the infrared.  Survey images of the
infrared continuum emission were presented by \citet[ $H$, $K$]{cam88}
and~\citet[ $J$, $H$, $K_{\text{s}}$]{ojh04} and showed diffuse emission, which
extends from the IRS1--3 cluster in a fan-shaped structure towards the
northeast and north, approximately tracing the optical \HII~region. The
northeast border of this NIR emitting region also appears very pronounced in
the continuum-subtracted H$_2$ 2.122~$\mu$m maps by~\cite{dav98}, possibly
tracing the illuminated surfaces of nearby molecular clouds or the inner walls
of a vast outflow cavity.  Furthermore, \citet{dav98} discovered two
bowshock-shaped structures, centered roughly on the IRS1--3 cluster and
orientated again along the northwest--southeast direction (PA~$\sim
-30^{\circ}$) in H$_2$~$2.122~\mu$m. With imaging at arcsecond resolution and
the use of several spectral filters ($J$, $H$, $K$, \FFeII~$1.65~\mu$m,
Br$\gamma$~2.165~$\mu$m, H$_2$~$2.122~\mu$m, and 3.29~$\mu$m), \cite{blo98}
attempted to identify the source and mechanism of the outflow.  Based on a
cometary-shaped morphology in the \FFeII~line images and shell-like rings
observed in the $J$, $H$, and $K$-bands, these authors propose a stellar wind
bowshock model in which the motion of IRS2 relative to the molecular cloud
produces the diffuse NIR emission within the vicinity of the IRS1--3 cluster.

The first $K$-band speckle images, taken with the 3.5~m-telescope on Calar Alto
were presented by~\cite{alv04} and showed substructure in the vicinity of
IRS1; namely, two strong blobs ($A$, PA~$\sim -45^\circ$; $B$, PA~$\sim
-70^\circ$), a diffuse emission feature ($C$, PA~$\sim 0^\circ$) as well as
several faint point-like sources ($a$-$f$).

%

\section{Observations}

\subsection{Bispectrum speckle interferometry}

\begin{figure*}[tbph]
  \centering
  \begin{tabular}{cc}
    \begin{minipage}{12cm}
      \centering
      \includegraphics[height=24cm]{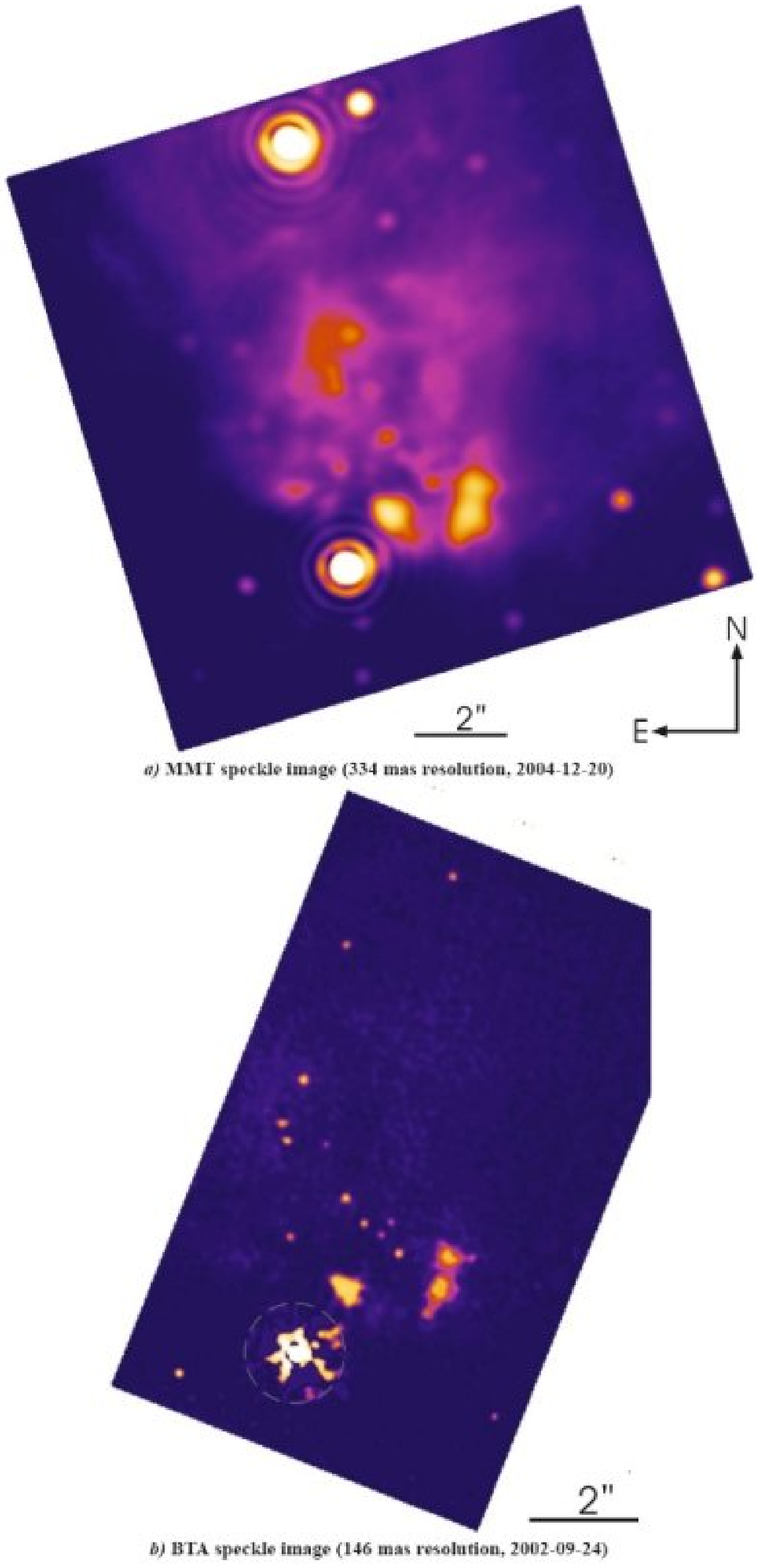}\\   
    \end{minipage}
    &
    \begin{minipage}{5cm}
      \caption{Bispectrum speckle images ($K'$-band) reconstructed from data taken
        with $a)$ the 6.5m MMT and $b)$ the 6m BTA telescope.  To show the weak emission features, the
        intensity of IRS1 was clipped to 2\% of the total flux.  Within the
        high-resolution image $(b)$, speckle-noise artifacts appear around
        IRS1 (marked with a circle). These weak features represent small
        distortions of the point-spread-function (PSF) on the 1\%-level and do
        not influence the reliability of the identification of point sources
        within the image. The absolute coordinates of IRS1 are
        $\alpha=\rab{23}{13}{45}{35}$ and
        $\delta=\decb{+61}{28}{10}{84}$~(J2000, determined from 2MASS,
        accuracy $\sim 0\farcs5$).}
      \label{fig:speckleimg}
    \end{minipage}
  \end{tabular}
\end{figure*}

The first set of observations was performed on 2002-09-24 using the 6.0~m
BTA (Big Telescope Alt-azimuthal) telescope of the Special
Astrophysical Observatory located on Mt. Pastukhov in Russia. Additional data
were gathered 2004-12-20 with the MMT (Multiple Mirror Telescope) on
Mt. Hopkins in Arizona, which harbors a 6.5~m primary mirror. As detector, we
used at both telescopes one 512$\times$512 pixel quadrant of
the Rockwell HAWAII array in our speckle camera.
All observations were carried out using a \textit{K'}-band filter centered on the
wavelength 2.12~$\mu$m with a bandwidth of 0.21~$\mu$m.
During the BTA observation run, we recorded 420 speckle interferograms on
NGC~7538~IRS1 and 400 interferograms on the unresolved star BSD
19-901 in order to compensate for the atmospheric speckle transfer function.  
The speckle interferograms of both objects were taken with an exposure time of
360~ms per frame.
For the MMT observations, the star 2MASS\,23134580+6124049 was used for the
calibration and 120 (200) frames were recorded on the target (calibrator) with
an 800~ms exposure time.
The modulus of the Fourier transform of the object (visibility) was obtained
with the speckle interferometry method~\citep{lab70}.  For image
reconstruction we used the bispectrum speckle interferometry method
(\citealt{wei77}, \citealt{wei83}, \citealt{loh83}, \citealt{hof86}).
With pixel sizes of 27.0~mas (BTA) and 28.7~mas (MMT) on the sky, the
reconstructed images possess fields of views of $13\farcs8$ (BTA) and
$13\farcs1$ (MMT), respectively.

We found that the BTA data allows the
highest spatial resolution (and is therefore perfectly suited for the
identification of point-sources within the field), whereas the image
reconstructed from the MMT data allows a high dynamic range in the diffuse
emission.  Therefore, we show the diffuse emission within an image of moderate
resolution (reconstructed from MMT data, see Figure~\ref{fig:speckleimg}$a$)
and perform point-source identifications within the higher resolution image
reconstructed from BTA data (Figure~\ref{fig:speckleimg}$b$).
In order to distinguish point-sources and diffuse structures reliably, we
reconstructed images of various resolutions (146~mas, 97~mas, 72~mas) and
carefully examined changes in the peak brightness of the detected features.
Whereas for point-sources the peak brightness increases systematically, it
stays constant or decreases for diffuse structures.

To perform an absolute calibration of the astrometry in our images, we
measured the position of IRS1 and IRS2 in the Two Micron All Sky Survey
(2MASS)~$K_s$~Atlas~images and use the determined absolute positions as
reference for our astrometry. We estimate that the accuracy reached in the
relative astrometry is $\sim 0\farcs1$. The absolute calibration introduces
further errors ($\sim 0\farcs2$).

\subsection{\textit{Spitzer}/IRAC Archive data}

\begin{figure*}[h]
  \centering
  \includegraphics[width=18cm]{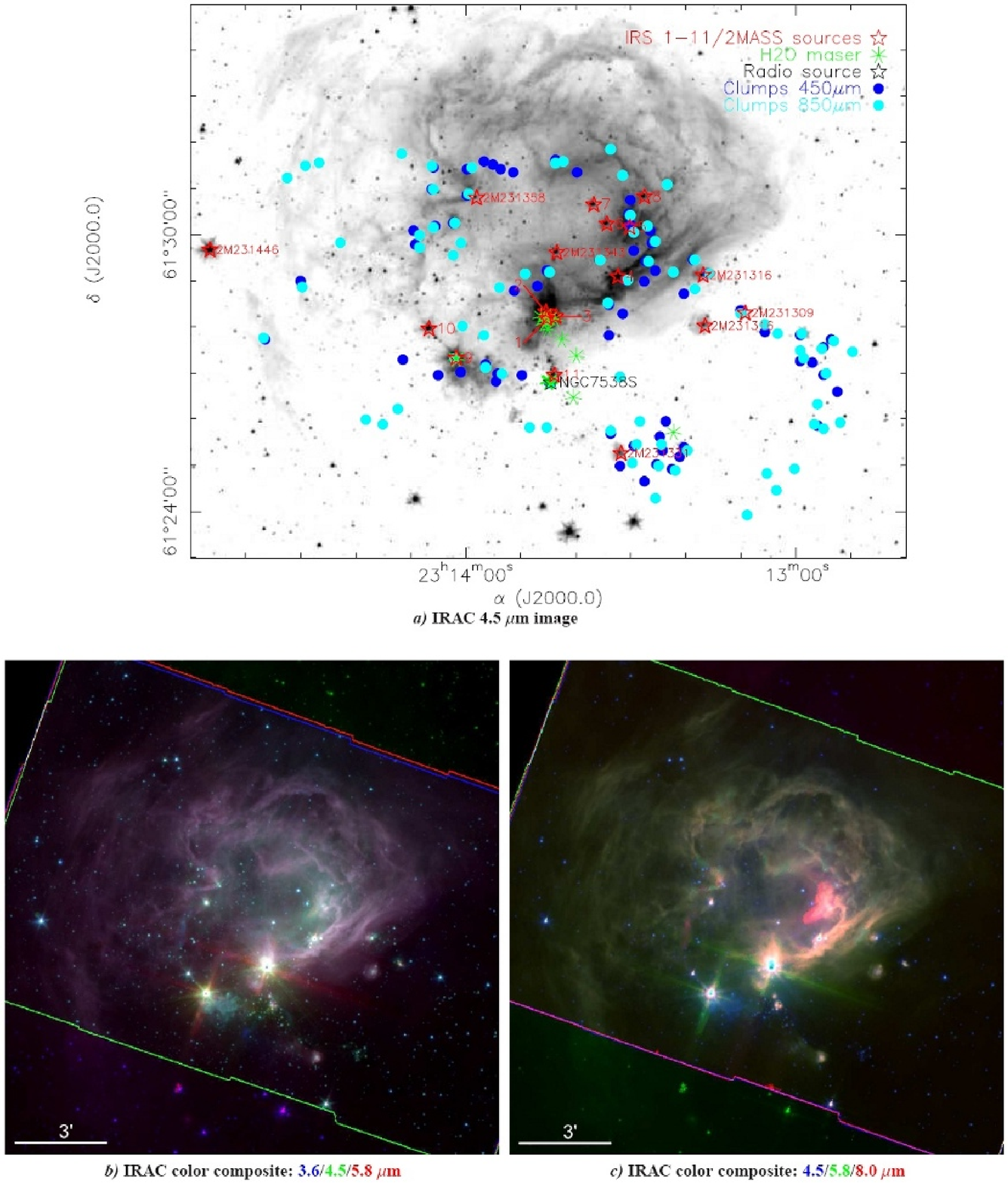}\\ 

  \caption{Figure \textit{a)} shows the Spitzer/IRAC 4.5~$\mu$m image with the position
    of the infrared sources (IRS1 to 11 and 2MASS sources) and H$_2$O masers
    marked. Furthermore, the position of the sub-millimeter (450~$\mu$m and
    850~$\mu$m) clumps reported by~\citet{rei05} are shown. The position of
    the 2MASS sources 2MASS\,23135808+6130484,
    2MASS\,23131660+6128017, 2MASS\,23134351+6129372, 2MASS\,23144651+6129397,
    2MASS\,23131691+6129076, 2MASS\,23130929+6128184, and
    2MASS\,23133184+6125161 are labeled explicitly.
    Figures \textit{b)} and \textit{c)} show color-composites produced with
    two triplets of the four IRAC bands.  The intensity of each image was scaled
    logarithmically. North is up and east to the left. }
  \label{fig:iracirs}
\end{figure*}

In order to relate our high-resolution images with the morphology of the
NGC~7538 molecular cloud at large scales, we examined archival 3.6,
4.5, 5.8, and 8.0~$\mu$m images (PI: G.~G.~Fazio), taken with the
Infrared Array Camera~\citep[IRAC, ][]{faz04} on the \textit{Spitzer} Space
  Telescope.  The four bands are recorded simultaneously using two InSb
(3.6~$\mu$m, 4.5~$\mu$m) and two Si:As (5.8~$\mu$m, 8.0~$\mu$m) detectors.
The central wavelengths and bandwidths of the IRAC bands~\citep{hor04} are
3.56~$\mu$m ($\Delta\lambda = 0.75$~$\mu$m),  
4.52~$\mu$m ($\Delta\lambda = 1.01$~$\mu$m), 
5.73~$\mu$m ($\Delta\lambda = 1.42$~$\mu$m), and
7.91~$\mu$m ($\Delta\lambda = 2.93$~$\mu$m).
Each image consists of $256\times256$~pixels, corresponding to a $\sim
5\arcmin \times 5\arcmin$ field-of-view on the sky.  The data used include
48~\textit{Spitzer} pointings taken on 2003 December 23 in the High Dynamic
Range~(HDR) mode.  In HDR~mode, for each pointing, images are taken with two
exposure times (0.6~s and 12~s) in order to record both bright and faint
structures.  However, the two brightest sources, IRS1 and IRS9, are saturated
even within the 0.6~s exposure.

We used the \textit{mopex} software (2005-09-05 version), released by the
Spitzer Science Center (SSC), to process both the long and short exposure
images.  Beside the basic calibration steps applied by the Basic Calibrated
Data (BCD) pipeline (S11.0.2), we performed Radhit detection, artifact
masking, and pointing refinement. Finally we generated a mosaic in
which the saturated pixels of the long exposure image were replaced by the
corresponding pixels of the 0.6~s exposure. The optical design of IRAC induces
a shift of $\sim 6\farcm 8$ between the 3.6/5.8~$\mu$m and 4.5/8.0~$\mu$m
pointings, leaving an overlap of $5 \farcm 1$ between all four bands.

In Figure~\ref{fig:iracirs}, color composites of the 3.6/4.5/5.8~$\mu$m and
4.5/5.8/8.0~$\mu$m band images are shown. 

The diffuse emission in three of the four IRAC bands is dominated by
Polycyclic Aromatic Hydrocarbons~\citep[PAHs,][]{chu04}, which trace the border
of regions excited by the UV photons from HMPOs particularly well.
Contributions are also expected from several vibrational levels of
H$_2$~\citep{smi05b}, atomic lines, CO vibrational bands, and thermal dust
grain emission. 

%

\section{Results}

\subsection{Bispectrum speckle interferometry: Small-scale structures around IRS1/2}

\subsubsection{IRS1 Airy disk elongation and diffuse emission}

\begin{figure*}[htbp]
  \centering
  \includegraphics[width=16cm]{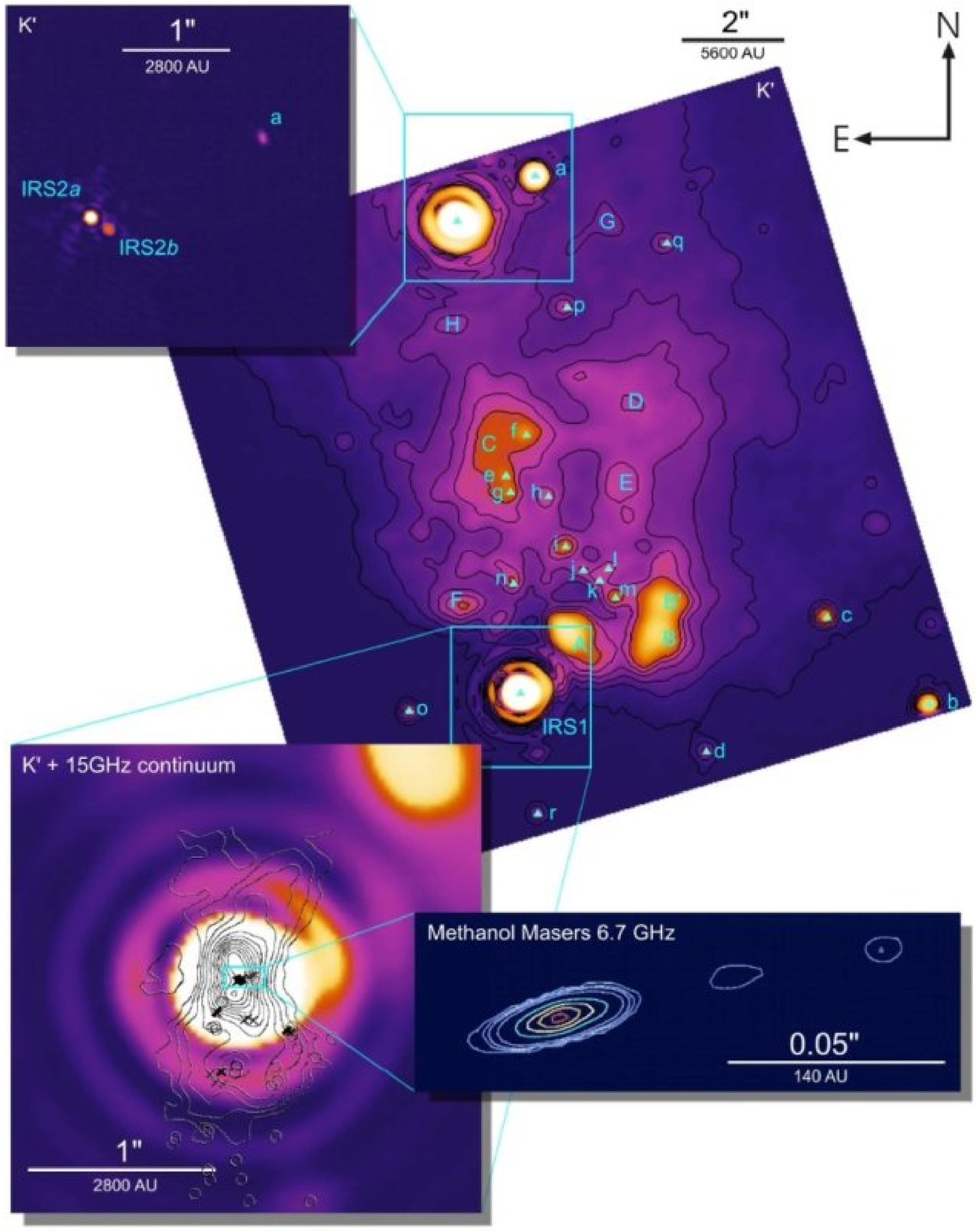}\\   
  \caption{Bispectrum speckle image with identified point sources (triangles) atop marked.
  The astrometry for the point-sources was performed using the high-resolution
  BTA image, whereas the image shown was reconstructed from MMT data.
  The contours trace 0.25\%, 0.5\%, 0.75\%, 1.0\%, 1.25\%, and 1.5\% of the
  peak intensity.
  The inset on the upper left shows a reconstruction of the vicinity of IRS2
  using a resolution of 80~mas (BTA data).  In the
  lower left, IRS1 is shown using a different color table, emphasizing the
  elongation of the IRS1 Airy disk (MMT data) overplotted with the 15 GHz
  radio continuum (the contours show -1, 1, 2.5, 5, 10, 20, ..., 90\% of the
  peak flux) and the position of the OH (circles) and methanol (crosses)
  masers (image from~\citealt{hut03} using data from~\citealt{gau95}).  In the
  lower right we show the integrated brightness of the methanol masers as
  presented by~\citet[][ contour levels of 1, 3, 5, 10, 30, 50, 70, and 90\%
  of the peak flux density are shown]{pes04}.}
  \label{fig:speckleimglabel}
\end{figure*}

In our speckle images, the Airy disk of IRS1 itself appears asymmetric, being
more extended towards the northwest direction (PA $\sim -70^\circ$, see
Figure~\ref{fig:speckleimg}$b$ and inset in the lower left of
Figure~\ref{fig:speckleimglabel}). In the same direction 
(PA $\sim-60^\circ$), we find two strong blobs ($A$, $B+B'$) of diffuse
emission at separations of $\sim\!1''$ and $2''$. These blobs and additional diffuse
emission seem to form a conical (fan-shaped) region with a $90^\circ$ opening
angle extending from IRS1 towards the northwest.

\begin{figure}[tbp]
  \centering
  \includegraphics[width=9cm]{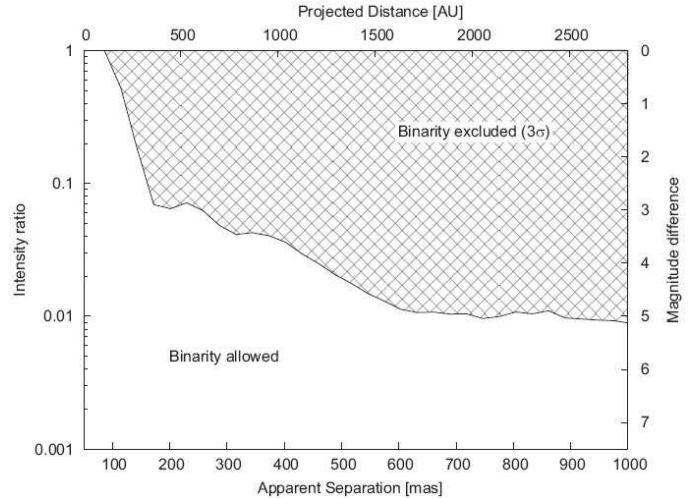}\\
  \caption{By measuring the speckle noise around the PSF of IRS1,
we can rule out binarity of IRS1 on a $3\sigma$-level as a function of
apparent separation and intensity ratio.}
  \label{fig:IRS1companionlimits}
\end{figure}

\begin{table*}
  \caption{NGC~7538~IRS1 outflow directions reported for various tracers. }
  \centering
  \begin{tabular}{l l c c c r c l}
    \hline\hline
    Tracer & Structure & Beam Size & Scale     & PA of              & Dynamical     & Ref. & Comments\\
           &           & [\arcsec] & [\arcsec] & outflow              & Age$^f$       &      & (Velocities in [\mbox{km~s$^{-1}$}])\\
           &           &           &           & direction [$^\circ$] & [$10^3$~yrs]  &      & \\
    \hline
    Methanol masers &                              & 0.03$^c$ &           & +19$^{a,d}$  &  --        &      [8]  & see~Fig.~\ref{fig:speckleimglabel}\\
    1.0~cm cont. & inner core $\lesssim 0\farcs5$  & 0.13  & 0.4          & +0$^{a}$     &  $ < 0.03$   &      [9]  & $t_e=0.15$~yrs$^e$\\
    1.3~cm cont. & inner core $\lesssim 0\farcs4$  & 0.11  & 0.4          & +0$^{a}$     &  $ < 0.02$   &      [5]  & \\
    6.0~cm cont. & inner core $\lesssim 0\farcs75$ & 0.35  & 0.4          & +0           &  $ < 0.04$   &      [1]  & \\
    1.0~cm cont. & outer core $\gtrsim 0\farcs5$   & 0.13  & 1.0          & -25$^{a}$    &  $ > 0.03$   &      [9]  & $t_e=0.3$~yrs$^e$\\
    1.3~cm cont. & outer core $\gtrsim 0\farcs4$   & 0.11  & 1.0          & -25$^{a}$    &  $ > 0.02$   &      [5]  & \\
    6.0~cm cont. & outer core $\gtrsim 0\farcs75$  & 0.35  & 1.0          & -15 ... -20  &  $ > 0.04$   &      [1]  & \\
    6.0~cm cont. & outer core $\gtrsim 0\farcs5$   & 0.4   & 1.0          & -25$^{a}$    &  $ > 0.03$   &      [9]  & \\
    MIR $11.7\mu$m        & IRS1~Elongation        & 0.43  & 3            &  -45         &    $0.16$    & [10] & see~Fig.~\ref{fig:mosaic}$k$\\
    MIR $18.3\mu$m        & IRS1~Elongation        & 0.54  & 4            &  -45         &    $0.16$    & [10] & see~Fig.~\ref{fig:mosaic}$l$\\
    NIR \textit{K'}-band      & IRS1~Elongation        & 0.3   & 0.6          & -78          &    $0.03$    & --   & see~Fig.~\ref{fig:speckleimglabel}\\
    NIR \textit{K'}-band      & feature~A              & 0.3   & 1.6          & -65          &    $0.08$    & --   & see~Fig.~\ref{fig:speckleimglabel}\\
    NIR \textit{K'}-band      & feature~F              & 0.3   & 2.1          & -33          &    $0.11$    & --   & see~Fig.~\ref{fig:speckleimglabel}\\
    NIR \textit{K'}-band      & feature~B              & 0.3   & 3.0          & -39          &    $0.16$    & --   & see~Fig.~\ref{fig:speckleimglabel}\\
    NIR \textit{K'}-band      & feature~B'             & 0.3   & 3.3          & -57          &    $0.18$    & --   & see~Fig.~\ref{fig:speckleimglabel}\\
    NIR \textit{K'}-band      & feature~C              & 0.3   & 4.8          & +6           &    $0.25$    & --   & see~Fig.~\ref{fig:speckleimglabel}\\
    NIR \textit{K'}-band      & feature~E              & 0.3   & 6.2          & -20          &    $0.33$    & --   & see~Fig.~\ref{fig:speckleimglabel}\\
    NIR \textit{K'}-band      & feature~D              & 0.3   & 7.4          & +10          &    $0.39$    & --   & see~Fig.~\ref{fig:speckleimglabel}\\
    NIR \textit{K'}-band      & eastern wall           & 0.3   & --           & +25          &   --         & --   & see~Fig.~\ref{fig:speckleimglabel}\\
    NIR \textit{K'}-band      & western wall           & 0.3   & --           & -65          &   --         & --   & see~Fig.~\ref{fig:speckleimglabel}\\
    \FFeII~$1.65~\mu$m &                        & 1        & 15$^a$       & N-S$^{a}$    &   $0.8$    & [7]  & around IRS2; see~Fig.~\ref{fig:mosaic}$i$\\
    H$_2$        & northern bow                 &          & 30$^a$       & N-S          &   $1.5$    & [6]  & \\
    H$_2$        &                              & 1        & 27$^a$       & -25$^{a}$    &   $1.4$    & [7]  & shell-like structure; see~Fig.~\ref{fig:mosaic}$i$\\
    H$_2$        & southern bow                 &          & 45$^a$       & 155$^{a}$    &   $2.3$    & [6]  & see~Fig.~\ref{fig:mosaic}$h$\\
    IRAC bands   & southern bow                 & 1        & 40           & 145          &   $2.2$    & --   & see~Fig.~\ref{fig:mosaic}$a$...$e$\\
    CO           & low velocities               & 7        & 5$^{a,b}$    & E-W          &   $0.9$    & [3]  & $-11<\tilde{V}_b<-6$; $2<\tilde{V}_r<9$; see~Fig.~\ref{fig:mosaic}$g$\\ 
    CO           & high velocities              & 7        & 15$^{b}$     & -35          &   $15$     & [3]  & $-17<\tilde{V}_b<-11$; $9<\tilde{V}_r<15$; see~Fig.~\ref{fig:mosaic}$g$\\ 
    CO           &                              & 34       & 18$^{b}$     & -50$^{a}$    &   $15$     & [2]  & $-24<\tilde{V}_b<-8$; $9<\tilde{V}_r<22$\\
    CO           &                              & 16       & 13$^{a,b}$   & -40          &   $14$     & [4]  & $-14<\tilde{V}_b<-9$; $11<\tilde{V}_r<16$; see~Fig.~\ref{fig:mosaic}$f$\\
    CO           &                              & 45       & 12$^{a,b}$   & -50$^{a}$    &   $10$     & [3]  & $-24<\tilde{V}_b<-8$; $9<\tilde{V}_r<22$\\
    \hline
  \end{tabular}

  \begin{flushleft}
    \hspace{5mm}Note~--~$\tilde{V}_r$ and $\tilde{V}_b$ are measured relative to the velocity of methanol maser feature A ($\tilde{V} = V - 56.25$~\mbox{km~s$^{-1}$})\\
    \hspace{5mm}$^{a}$ Estimated from figures presented within the reference paper; therefore, with limited accuracy.\\
    \hspace{5mm}$^{b}$ The half-separation between the red- and blue-shifted CO lobe is given.\\
    \hspace{5mm}$^{c}$ For VLBI observations, we give the estimated error on the absolute position of the measured maser spots.\\
    \hspace{5mm}$^{d}$ The expected outflow direction is given; i.e., perpendicular to the measured orientation of maser feature A.\\
    \hspace{5mm}$^{e}$ Electron recombination time given in the reference paper.\\
    \hspace{5mm}$^{f}$ Assuming an outflow velocity of 250~\mbox{km~s$^{-1}$}, which was measured by~\citet{gau95} within the H66$\alpha$ recombination line. For the CO emission, we also use the measured CO outflow velocity and provide the corresponding dynamical age in brackets.  Since all velocities are measured along LOS, this timescale can only provide upper limits.\\
    References: [1]~\citet{cam84b}; [2]~\citet{fis85}; [3]~\citet{sco86}; [4]~\citet{kam89}; [5]~\citet{gau95}; [6]~\citet{dav98}; [7]~\citet{blo98}; [8]~\citet{min00}; [9]~\citet{her04}; [10]~\citet{deb05}
  \end{flushleft}
  \label{tab:tracers}
\end{table*}

\begin{figure*}[htbp]
  \centering
  \begin{tabular}{cc}
    \begin{minipage}{5cm}
      \caption{Illustration showing the outflow directions in the various
        tracers.  The CO contours by~\citet[ red and blue]{kam89} are overlaid on the
        H$_2$ map (greyscale) by~\cite{dav98}. The orientation of the
        conjectural methanol maser disk (green), the fan-shaped structure
        detected in our \textit{K'}-band image (orange), and the averaged 
        direction of H$_2$~(red arcs) are shown schematically. The arrows
        indicate the direction prependicular to the alignment of the methanol
        masers (green), the orientation of the inner ($<0.5\arcsec$) and
        outer ($>0.5\arcsec$) core detected in the 1.0, 1.3, and 6.0~cm radio 
        continuum (white), and the direction along which the IRS1 Airy disk
        was found to be elongated (MIR:~\citealt{deb05}; NIR: this paper). } 
      \label{fig:illustration}
    \end{minipage}
    &
    \begin{minipage}{11cm}
      \centering
      \includegraphics[width=11cm]{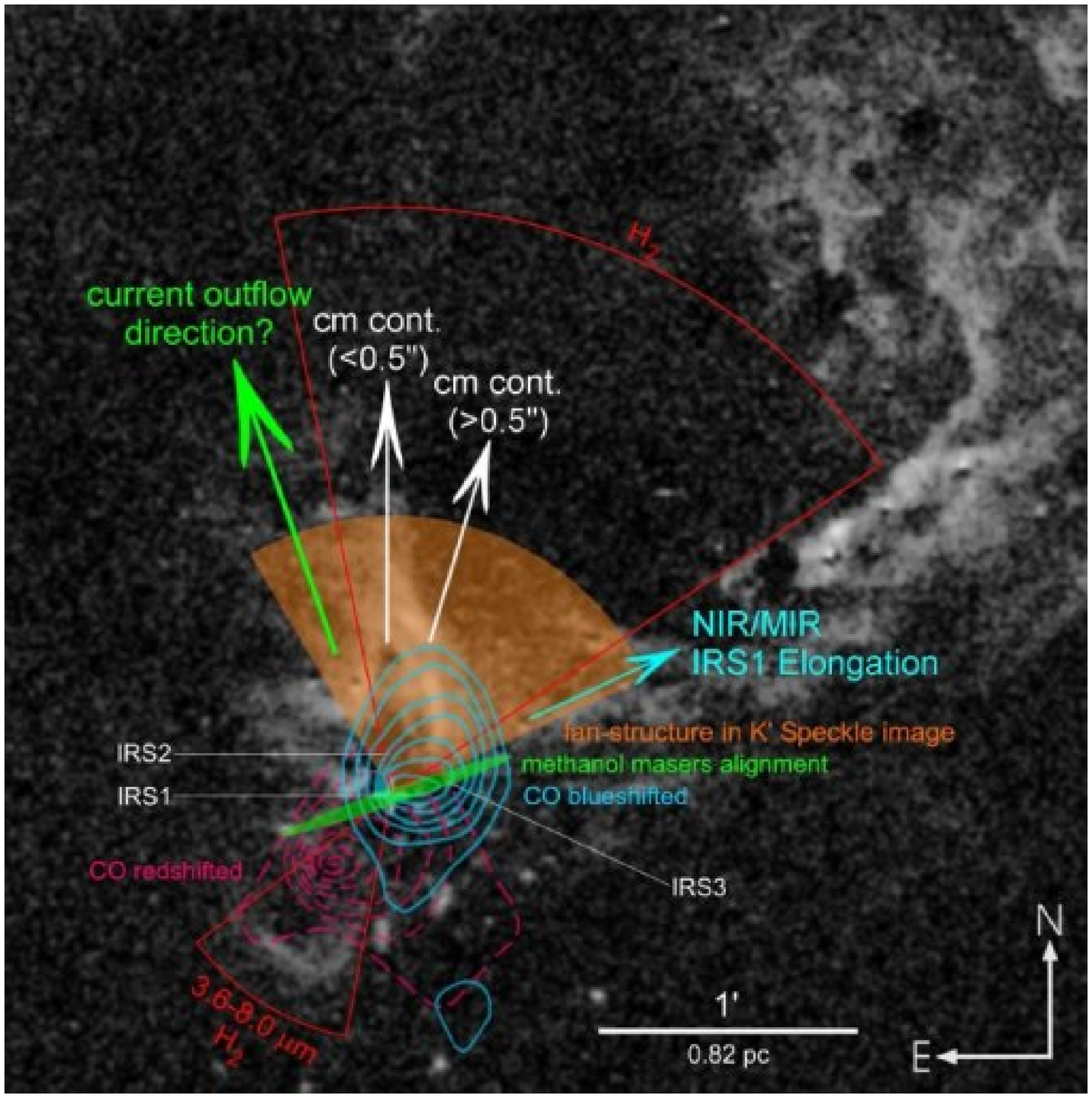}
    \end{minipage}
  \end{tabular}
\end{figure*}

A careful examination of the power spectrum of IRS1 has shown that the detected
asymmetry of IRS1 is not caused by a companion, but seems to represent
diffuse emission.  Therefore, we can rule out a close binary system of
similar-brightness components down to the diffraction limit of $\sim 70$~mas.
For the case of a binary system with components of significantly different
brightnesses, we can put upper limits on the brightness of the hypothetical
companion as a function of the projected binary separation (see
Figure~\ref{fig:IRS1companionlimits}). 

The PA of the elongation of the Airy disk is similar to the PA of \textit{K'}
blobs $A$, $B$, and $B'$.
Another strong feature ($C$) can be seen towards PA$\sim 0^\circ$.
The blobs seem to be connected by a bridge of diffuse emission extending
from feature $B$ to $C$. Overall, the diffuse emission seems to form a
fan-shaped region which is extending from IRS1 towards the northwest with an
opening angle of nearly $90^\circ$ .  We identified some further features and
list their position angles and separations in Table~\ref{tab:tracers}.
The directions, which were reported for various outflow tracers, are also
listed in this table and illustrated in Figure~\ref{fig:illustration}.

Our features $A$, $B+B'$, and $C$ appear to coincide with the features $A$, $B$,
and $C$ identified by~\citet{alv04}.  A comparison suggests that features
$A+B+B'$, $C$, $D$ correspond to the peaks $1W$, $1N$, and $1NW$ in~\citet{tam91}.

\subsubsection{Binarity of NGC~7538~IRS2} \label{cha:IRS2binarity}

IRS2 appears resolved as a close binary system.
Using an image reconstructed from BTA~data with a spatial resolution of
80~mas~(see inset in the upper-left corner of
Figure~\ref{fig:speckleimglabel}), we determined the separation to be 195~mas and
found a PA of $-123^\circ$ for the 1\fracmag9 fainter companion
(2002-09-24).
We designate the brighter component in the \textit{K'}-band as IRS2$a$
and the fainter as IRS2$b$.

\subsubsection{Detection of fainter cluster members}\label{cha:lowmassstars}

\begin{table*}
  \caption{Point sources identified in our speckle images.  For details about IRS1~and
    the binary system~IRS2$a$/$b$, we refer to the text.  We identify components $a$
    to $f$ with the stars already discovered in the image by~\cite{alv04}. }
  \label{tab:pointsources}
  \centering
  \begin{tabular}{c c c c l}
    \hline\hline
    Name & RA~(J2000)$^a$ & DEC~(J2000)$^a$ & \textit{K'} Magnitude$^b$ & Comment\\
    \hline
       a & \rahm{23}{13}\ras{45}{33} & \decdm{61}{28}\decs{21}{02}& 10\fracmag94 & \\
       b & \hspace{11mm}\ras{44}{81} & \hspace{10mm}\decs{10}{63} & 11\fracmag29 & \\
       c & \hspace{11mm}\ras{44}{95} & \hspace{10mm}\decs{12}{33} & 11\fracmag46 & \\
       d & \hspace{11mm}\ras{45}{11} & \hspace{10mm}\decs{09}{69} & 11\fracmag73 & \\
       e & \hspace{11mm}\ras{45}{37} & \hspace{10mm}\decs{15}{12} & 11\fracmag77 & embedded in feature $C$\\
       f & \hspace{11mm}\ras{45}{34} & \hspace{10mm}\decs{15}{93} & 11\fracmag73 & embedded in feature $C$\\
       g & \hspace{11mm}\ras{45}{36} & \hspace{10mm}\decs{14}{80} & 11\fracmag73 & embedded in feature $C$\\
       h & \hspace{11mm}\ras{45}{31} & \hspace{10mm}\decs{14}{72} & 11\fracmag86 & \\
       i & \hspace{11mm}\ras{45}{29} & \hspace{10mm}\decs{13}{74} & 11\fracmag71 & \\
       j & \hspace{11mm}\ras{45}{27} & \hspace{10mm}\decs{13}{25} & 11\fracmag73 & \\
       k & \hspace{11mm}\ras{45}{25} & \hspace{10mm}\decs{13}{05} & 11\fracmag81 & \\
       l & \hspace{11mm}\ras{45}{24} & \hspace{10mm}\decs{13}{28} & 11\fracmag83 & \\
       m & \hspace{11mm}\ras{45}{22} & \hspace{10mm}\decs{12}{71} & 11\fracmag71 & \\
       n & \hspace{11mm}\ras{45}{36} & \hspace{10mm}\decs{12}{99} & 11\fracmag75 & \\
       o & \hspace{11mm}\ras{45}{50} & \hspace{10mm}\decs{10}{50} & 11\fracmag60 & \\
       p & \hspace{11mm}\ras{45}{29} & \hspace{10mm}\decs{18}{42} & 11\fracmag73 & \\
       q & \hspace{11mm}\ras{45}{16} & \hspace{10mm}\decs{19}{69} & 11\fracmag77 & \\
       r & \hspace{11mm}\ras{45}{33} & \hspace{10mm}\decs{08}{48} & 11\fracmag71 & \\
    \hline
  \end{tabular}
  \begin{flushleft}
    \hspace{5mm}$^{a}$ For the astrometry, the relative errors are of the
    order of 0\farcs1.  The absolute calibration using the reference
    position of IRS1 in 2MASS introduces further errors (0\farcs2).\\
    \hspace{5mm}$^{b}$ The photometry was done relative to IRS1 with an
    uncertainty of 0\fracmag3.  For the conversion to absolute photometry, we
    assumed a IRS1 magnitude of 8\fracmag9~\citep{ojh04}.\\
  \end{flushleft}
\end{table*}

Besides IRS1 and IRS2$a/b$, we were able to identify 18~additional fainter
point-like sources ($a$-$r$) within the BTA image, whose positions and
\textit{K'}-band magnitudes are listed in Table~\ref{tab:pointsources}. 
 
In order to test whether these sources are physically related to NGC~7538, one
can compare the stellar number density for the brightness range 11\fracmag0 to
12\fracmag0 in our speckle image, 
$N_{\mathrm{Speckle}}=18/128~\mathrm{arcsec}^{2}\approx 2.1\times10^6/\mathrm{deg}^{2}$,
with the number expected from the cumulative \textit{K}-band luminosity
function (KLF) of the NGC~7538 field\footnote{The \textit{K}-band luminosity
  function by~\citet{bal04} for the whole NGC~7538 region, corrected
  with the on-cluster KLF, and cumulated for the magnitude range 11\fracmag0
  to 12\fracmag0 was used.}
$N_{\mathrm{field}}\approx 1.8\times10^3/\mathrm{deg}^{2}$.
Although these number densities were obtained with different spatial
resolution, the clear over-density of stars in our speckle image is
significant and we conclude that most of the detected stars are likely members
of the NGC~7538 star forming region. 
When using the KLF for the IRS~1-3
region instead of the whole NGC~7538 field, the stellar over-density in our
speckle image becomes even more evident ($N_{\mathrm{IRS1-3\,cluster}}\approx
1.4\times10^3/\mathrm{deg}^{2}$).
Since these stars are about 5 to 6 magnitudes fainter than IRS1, they
are likely to be part of the associated intermediate mass stellar
population.

The arrangement of the stars within the fan-shaped nebula does not appear to
be random, but follows the \S-structure of the diffuse blobs (see
Figure~\ref{fig:speckleimglabel}). Most remarkable, more than half of the
stars seem to be aligned in a chain reaching from feature $B$ to $C$
(PA~$\sim 45^\circ$). Within the diffuse blobs close to IRS1 ($A$, $B$, $B'$),
no stars were found, whereas embedded in blob $C$, three stars could be
detected.

%

\subsection{\textit{Spitzer}/IRAC: Morphology at large spatial scales}

Imaging of NGC~7538 at optical wavelengths showed that diffuse emission
can mainly be found in the vicinity of IRS5~\citep{lyn86}.  At near-infrared (NIR)
wavelengths~\citep{ojh04}, a diffuse structure can be found extending from the
IRS1-3 cluster towards the northwest with the strongest emission around IRS5.

The Spitzer/IRAC images reveal a more complex, bubble-like structure (see
Figure~\ref{fig:iracirs}$b$, $c$), whose western border is formed by a
pronounced ridge-like filament connecting IRS1-3 with IRS4 and reaching up to
IRS5 (see Figure~\ref{fig:iracirs}$a$). 
At the western border of the bubble a wide conical structure is located, with
a vertex on 2MASS\,23135808+6130484.  Another conical structure can be detected close to
the northern border of the bubble.   Several other outflow structures can be
found in the IRAC image; most noteworthy, the unidirectional reflection nebula
around 2MASS\,23144651+6129397, 2MASS\,23131691+6129076, and
2MASS\,23130929+6128184 (see Figure~\ref{fig:iracirs}$a$). The sources
2MASS\,23131660+6128017 and 2MASS\,23133184+6125161 appear to be embedded in a
shell-like cloud.

Besides the position of the strongest near-infrared sources,
Figure~\ref{fig:iracirs} shows also the position of the submillimeter
(450~$\mu$m, 850~$\mu$m) clumps reported by~\citet{rei05}.  These clumps trace
the filaments and knots of the bubble, which can be seen in the IRAC
images, very well.  Besides this, the submillimeter clumps suggest another bubble-like
structure to the southwest of IRS4 (see also the images in~\citealt{rei05}).
This bubble seems to be invisible at near- and mid-infrared wavelengths,
although several NIR sources are located on its border
(2MASS\,23130929+6128184, 2MASS\,23133184+6125161).

As already pointed out by~\citet{rei05}, it is interesting to compare the
position of the detected H$_2$O masers with the position of the centers of
high-mass star formation in the region and to find agreement in many cases
(IRS1--3, IRS9, NGC~7538S).  However, as can be seen in
Figure~\ref{fig:iracirs}, for four locations of H$_2$O masers, no MIR
counterpart can be found in the IRAC images (the detection limits for point
sources in the four IRAC bands are roughly 3.6, 5.3, 31, and 34~$\mu$Jy for
the IRAC bands at 3.6, 4.5, 5.8, and 8~$\mu$m assuming medium sky
background).

%

\section{Discussion}

\subsection{Nature of the observed \textit{K'}-band emission}

\begin{figure*}[h]
  \begin{center}
   \includegraphics[height=22cm]{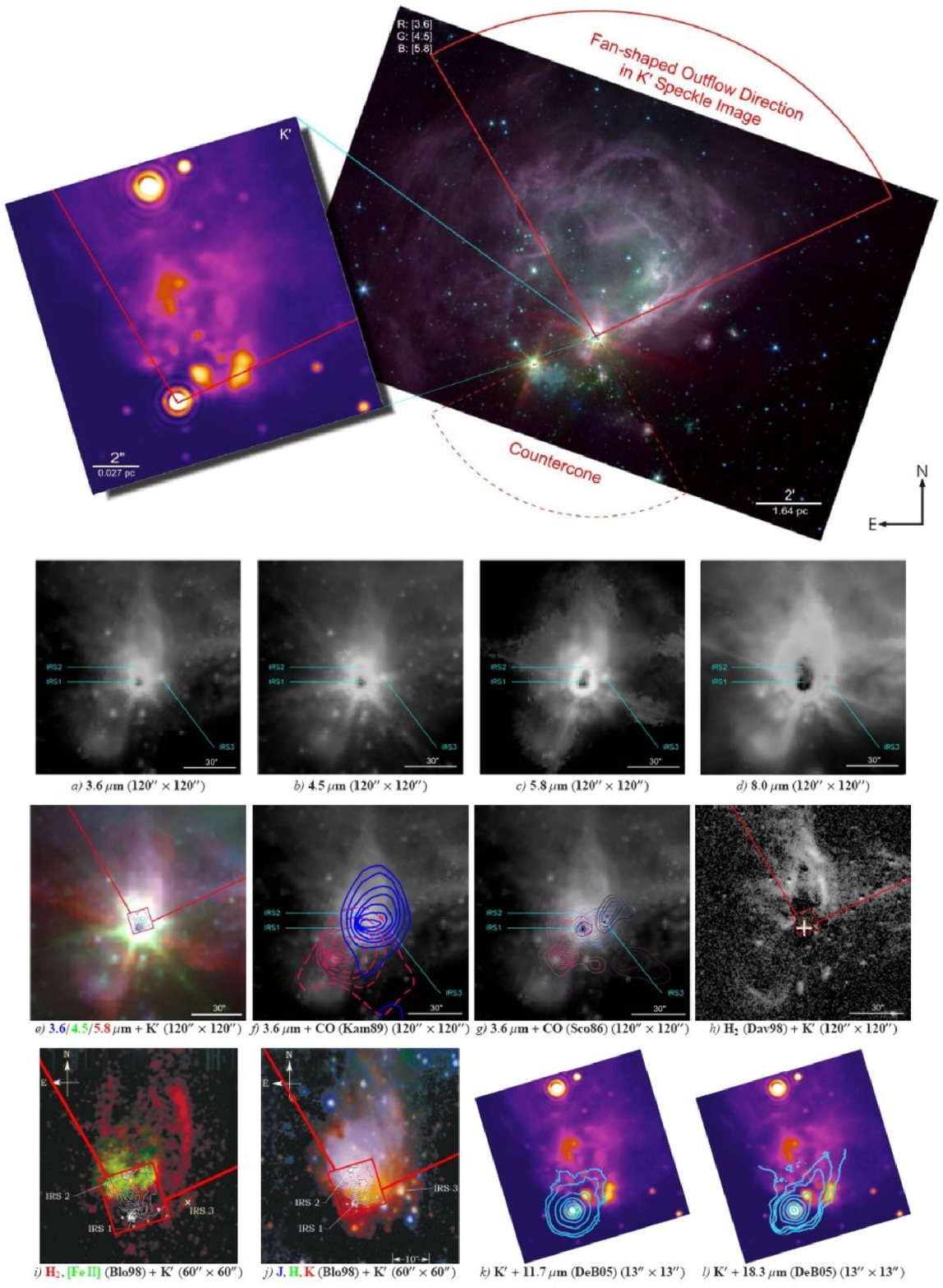}\\[3mm]
  \end{center}
  \caption{Mosaic showing the IRS1--3 cluster at various wavelengths. Beside
    the speckle $K'$-band image (also marked as red box) and IRAC images, data
    from~\citet[][ CO]{sco86}, \citet[][ CO]{kam89}, \citet[][ $J$, $H$, $K$, H$_2$,
    \FFeII]{blo98}, \citet[][ H$_2$]{dav98}, and \citet[][ $11.7~\mu$m,
    $18.3~\mu$m]{deb05} was incorporated.}
  \label{fig:mosaic}
\end{figure*}

In the wavelength range of the \textit{K'}-band filter
($\lambda_0=2.12$~$\mu$m, $\Delta\lambda=0.21$~$\mu$m), we record not
only continuum radiation (e.g.~scattered light, thermal dust emission, stellar
continuum emission), but also line emission (e.g.~H$_2$).  However,
both~\citet[ see Figure~\ref{fig:mosaic}]{blo98} and~\citet{dav98} did not
detect significant amounts of H$_2$~emission around IRS1. Furthermore, deep
optical imaging by \citet{els82} and \citet{cam88} reveal a weak optical source
offset $2\farcs2$ north of the radio source IRS1. The latter authors argue that
the strong extinction ($A_V=13$) derived for IRS1 makes it highly unlikely that
this optical emission is connected to IRS1 itself but that it most likely
represents scattered light. The measured offset suggests that the faint optical
source should be associated with blobs $A$ and $B$ in our images, making
scattering the most likely radiation mechanism for the detected
\textit{K'}-band emission. This conclusion is also supported by polarization
measurements~\citep{dyc79} which show a strong polarization of the 2~$\mu$m
emission, tracing either scattered light or light transmitted through aligned
grains. Henceforth we presume continuum to be the most important contributor to
the detected emission.

\subsection{Methanol maser feature A: Protostellar disk or outflow?}

We note that the 2MASS position of IRS1
($\alpha~=~\rab{23}{13}{45}{35}$,~$\delta~=~\decb{+61}{28}{10}{8}$,~J2000) and
the position of the methanol maser feature A
($\alpha~=~\rab{23}{13}{45}{364}$, $\delta~=~\decb{+61}{28}{10}{55}$, J2000)
reported by~\citet{min00} 
coincide within the errors\footnote{The astrometric accuracy of the 2MASS
catalogue was reported to be $\sim 0\farcs15$ (see
http://ipac.caltech.edu/2mass/releases/allsky/doc/expl-sup.html).}. Therefore,
the methanol masers and the outflow driving source are likely causually
connected, however a random coincidental alignment cannot be ruled out.\\

Since methanol masers can trace both protostellar disks and outflows, it is
not a~priori clear how the linear alignment of the methanol maser feature A
and the observed velocity gradient should be interpreted.  For IRS1, both
claims have been made \citep{pes04,deb05}.  However, detailed modeling has
provided strong quantitative support for the disk interpretation but is
still missing for the outflow interpretation.  Furthermore, a study
by~Pestallozi~et~al.~(in prep.) suggests that simple outflow
geometries cannot explain the observed properties of feature A.\\

A major difference between these two scenarios is the orientation of the disk
associated with the outflow driving source:  Whereas in the disk scenario the
methanol masers are lined up within the disk plane (PA~$\sim -62^\circ$), the
outflow scenario suggests an orientation of the disk plane perpendicular to the
maser alignment (PA~$\sim +28^\circ$). The observed asymmetry in our NIR
speckle images, as well as the elongation of the emission observed in the
11.7~and 18.3~$\mu$m images by~\cite{deb05}, can be explained within both
scenarios:\\[3mm]
\noindent
\textbf{Scenario A:}
  If maser feature A traces an \textbf{outflow
  cavity}, the detected asymmetry might simply reflect the innermost walls of
  this cavity (oriented northwest), whereas the southeastern cavity of a
  presumably bipolar outflow might be hidden due to inclination effects.\\[3mm]
\noindent
\textbf{Scenario B:}
  Alternatively, if the masers trace an edge-on \textbf{circumstellar disk},
  the asymmetry of the infrared emission could trace the western wall of an
  outflow cavity with a wide half-opening angle. The asymmetry cannot be
  attributed to the disk itself because the detection of stellar radiation
  scattered off the disk surface at such a large distance is highly unlikely.\\

For completeness, we also mention the interpretation by~\citet{kam89}, who
attributed the change between the direction observed in the UC\HII~region
(PA~$0^\circ$) and the high-velocity CO flow (PA~$-60^\circ$) to flow
deflection, either by large-scale magnetic fields or due to density
gradients. 

We proceed now to discuss both scenarios within an outflow-cavity model
(Sec.~\ref{cha:outflowcavity}) and a precessing jet model
(Sec.~\ref{cha:jetprecession}), incorporating the large amounts of evidence
collected by various authors over the last three decades.

\subsection{Scenario A: Outflow cavity model} \label{cha:outflowcavity}

Since the intensity of the diffuse emission in our images seems to decrease
with distance from IRS1 and the vertex of the fan-shaped region appears
centered on IRS1, we cannot support the hypothesis by~\citet{blo98}, who
identified IRS2 as the likely source of the diffuse NIR emission. Instead, the
observed fan-shaped region can be interpreted as a cavity that was formed by
outflow activity from IRS1. Because the walls of the fan-shaped structure are
well-defined, we can measure the opening angle of the proposed outflow cavity
from the eastern wall (PA~$25^\circ$) to blob $A$ (PA~$-65^\circ$), obtaining
a wide total opening angle of $90^\circ$. 

The unidirectional asymmetry of IRS1 in the NIR and MIR images (see
Figure~\ref{fig:mosaic}) is naturally explained in this context as scattered
light from the inner ($< 1500$~AU) walls of the cavity.  This scenario is also
consistent with the southeast--northwest orientation of the CO~outflow, aligned
roughly parallel to the methanol masers (PA~$\sim -62^\circ$).  Blobs $A$,
$B$, $B'$ are located within the same direction and might resemble either
clumps in the cavity or recent ejecta from the outflow.  The various blobs
might also indicate the presence of several outflows.

In order to resolve the misalignment of the radio-continuum core with respect to
the other outflow tracers, it was proposed that the radio-continuum emission
might arise from a photoevaporated disk wind \citep{lug04}.

However, as noted above, the methanol maser feature A lacks a quantitative
modeling up to now.

\subsection{Scenario B: Precessing jet model} \label{cha:jetprecession}

\begin{figure*}[htbp]
  \centering
  \includegraphics[width=18cm]{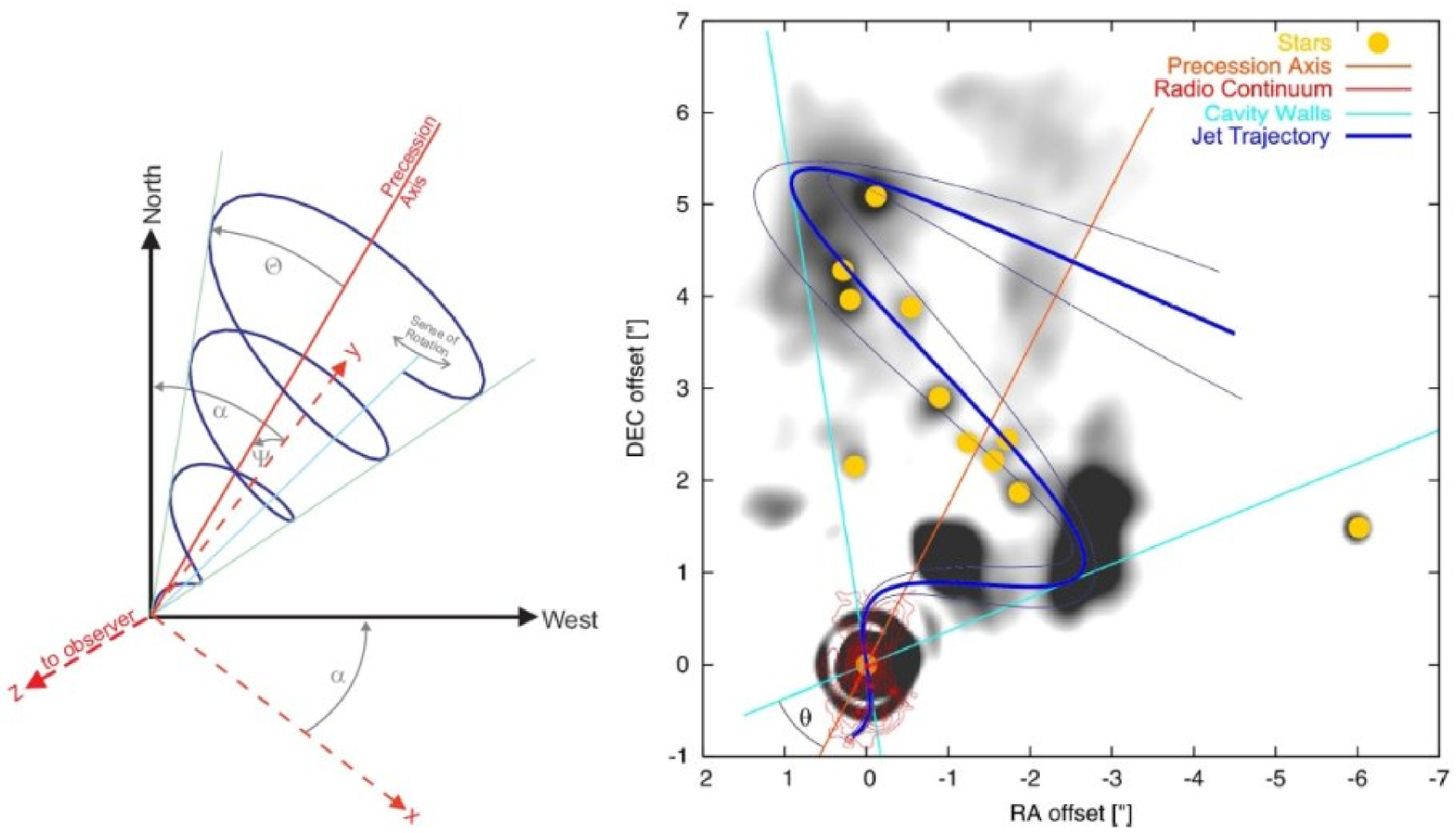}\\

  \caption{\textit{Left:} Illustration of the analytic precession model
    presented in Sec.~\ref{cha:anaprecmodel}.
    \textit{Right:} MMT speckle image overplotted with the trajectory of
    ejecta from a precessing outflow projected onto the plane of the sky
    (thick blue line) as described by the analytic precession model.
    For the counter-clockwise precession, the parameters
    $P_\text{prec}=280$~yrs, $\Theta=40^\circ$, $\alpha=-30^\circ$,
    $\psi=0^\circ$ (precession axis within the plane 
    of the sky), and $\phi_{0}=0^\circ$ were used.  In order to simulate a 
    finite collimation of the flow, we varied both $\alpha$ and $\psi$ by 
    $\pm 5^\circ$, yielding the trajectory given by the thin blue lines.
    The red contours show the 15 GHz radio continuum map
    by~\citealt[]{hut03} (using data by~\citealt{gau95}).
  }
  \label{fig:precession}
\end{figure*}

\subsubsection{Constraints from the methanol maser disk}
\label{cha:maserconstraints}

The circumstellar-disk modeling presented by~\citet{pes04} reproduces the
observational data for maser feature A in minute detail.  Assuming a central
mass of $30M_{\sun}$ and Keplerian rotation, this model confines the inner
($r_i \sim 290$~AU) and outer ($r_o \sim 750$~AU) radii of the disk (these
radii scale as $(M/30M_{\sun})^{-1/3}$ with the central mass $M$). The model
does not set strict constraints on the inclination and orientation of the disk
on the sky.

An uncertainty in the disk inclination arises from the assumption that
methanol is formed within a surface layer of the disk from photoevaporation of
H$_2$O. The midplane of the disk might therefore be inclined within certain
limits. A distinct inclination is suggested by the fact that the NIR/MIR
continuum emission, as well as the H$_2$ shock tracer line emission, appears
more pronounced towards the north than towards to south (see
Figures~\ref{fig:speckleimg} and~\ref{fig:mosaic}$h$). An inclination of the
northern outflow towards us is also indicated by CO outflow
observations~\citep[e.g.][]{kam89}, which show the blueshifted CO lobe of IRS1
towards the northwest (see Figures~\ref{fig:mosaic}$f$~\&~\ref{fig:mosaic}$g$).

The disk orientation on the sky can only be constrained by the maser
observations with a limited accuracy since the masers only trace a narrow
latitudinal arc of the disk, missing potential disk warping. Nevertheless, it
is still reasonable to identify the disk orientation with the linearly-aligned
feature A.

\subsubsection{Indications for disk and jet precession}\label{cha:precindications}

Assuming that the alignment of the masers is representative of the orientation
of the disk midplane (i.e. assuming disk warping is
negligible\footnote{Interestingly, the converse assumption (that disk warping
  is non-negligible) implies disk precession as well, as the jet would be
  launched perpendicular to the warped surface of the inner part of the disk.}),
it is evident that the direction perpendicular to the disk plane
(PA~$+19^\circ$, the expected outflow direction) is significantly misaligned
from the axes of the bipolar CO outflow and the NIR fan-shaped region
(PA~$\sim -20^\circ$; illustrated within Figure~\ref{fig:speckleimglabel}).
Also, the observed bending~\citep{cam84b, gau95} in the radio continuum could
indicate a change in the outflow direction. Whereas the inner core ($\lesssim
0\farcs5$) is orientated along PA~$\sim 0^\circ$, the outer core ($0\farcs5$
-- $1\farcs0$) bends slightly towards the west (PA~$\sim -25^\circ$).  This
might indicate that the outflow changed its direction by this amount within
the times needed by the jet to propagate the appropriate projected distances
($\sim25$~and $\sim50$~years). 

The bending detectable in the UC~\HII~region on scales of $\lesssim 1\farcs0$
seems to continue at larger scales within the morphology observed in
our speckle images, suggesting an \S-shaped fine-structure of the diffuse
emission extending from IRS1 initially towards the northwest and further out
towards north.  The blobs $A$, $B$, and $B'$ observed close to IRS1
(PA~$\sim-60^\circ$) might represent the most recent ejecta, whereas the
weak features which appear further away in our images ($C$, $D$)
might trace earlier epochs of the history of the outflow.

Based on these indications, we suggest a disk and jet precession model.
The fan-shaped diffuse emission in which the \S-structure is embedded can be
explained as scattered light from the walls of an outflow cavity, which was
cleared by the proposed wandering jet.

The western wall of this wide, carved-out outflow cavity might appear within
our NIR and the MIR images as an elongation of IRS1.  Since this elongation
extends mainly towards the northwest, there must be an additional reason why
the western wall of this cavity appears more prominent than the eastern wall. A
possible explanation might be shock excitation of the western wall, which would
cool through emission in shock tracer lines like H$_2$, which is contributing
to the recorded \textit{K'}-band.

Assuming the precession period derived in Sec.~\ref{cha:anaprecmodel}, the
outflow (which currently points towards PA~$\sim 19^\circ$) would have
excited the western wall of this cavity $\sim140$~years ago, which corresponds
roughly to the H$_2$ radiative cooling time.

The arrangement of the fainter cluster members embedded within the diffuse
emission can be understood in this context, too: Taking into account that IRS1
is still deeply embedded in its natal circumstellar cloud, the jet would have
cleared the envelope along its wandering path. The decreasing column density results in
lower extinction along the jet's path, revealing the fainter stars which
likely formed in the vicinity of IRS1.  The fainter stars might therefore be
detectable only in those regions where the precessing jet reduces the
extinction sufficiently. Within the blobs closest to IRS1, stars may be
undetectable because of either inclination effects or confusion with the
significantly higher surface brightness of blobs IRS1 $A$, $B$, and $B'$
(limiting the sensitivity to detect point sources), or because of the high
density of the outflowing material itself, providing intrinsic extinction.

The outflow tracers observed at rather large scales (CO, H$_2$, see
Figure~\ref{fig:mosaic}) are oriented roughly in the same direction as the NIR
fan-shaped structure.  
The CO channel maps by~\citet[ Figure~\ref{fig:mosaic}\textit{g}]{sco86}
suggest a change in the orientation of the CO outflow lobes for low and high
velocities.  Whereas the low velocity CO outflow is oriented along the
east-west direction, the high velocity lobes are oriented along PA~$\sim -35^\circ$.
As CO traces material swept-up by the outflow and has a relatively long cooling
time (of the order of $10^4$~yrs), the different orientations observed at
low- and high velocities are more difficult to interpret.

Finally, we speculate that the precession model might also explain why the
velocities of the methanol maser features B, C, D, and E are in the same range
as the velocities of the CO outflow \citep{deb05}, but show opposite signs for
the LOS velocity with respect to feature A (maser features B, C, D, and E are
blue-shifted, whereas the southern CO lobe is red-shifted).  Assuming
precession, the CO outflow would trace the average outflow direction around
the precession axis (with the southern axis oriented away from the observer),
whereas the methanol masers might trace clumps very close to the source, which
were excited more recently when the southern part of the outflow was pointing
towards the observer\footnote{This is consistent with the precession parameters determined
  in Sect.~\ref{cha:anaprecmodel}, where we find that the half-opening angle
  of the precession $\Theta$ is larger than the inclination of the precession
  axis with respect to the plane of the sky $\psi$.}.

In general, precession can explain the change in the flow orientation, but
potential alternative explanations include density gradients in the
surroundings of IRS1, the presence of multiple outflows, and flow deflection.

\subsubsection{Analytic precession model}
\label{cha:anaprecmodel}

In order to get a rough estimate for the precession parameters, we employ a
simple analytical model with constant radial outflow speed ${\mathrm v}$. On
the radial motion we superpose a precession with period $P_{\text{prec}}$,
leading to the wave number
\begin{equation}
  k = \frac{2\pi}{{\mathrm v} P_{\text{prec}}}\,;
\end{equation}
by the time that ejected material travels a distance $r$ from the source, the
direction of the jet axis changes by the angle $kr$. 

To describe the jet propagation in three dimensions we introduce a Cartesian
coordinate system centered on IRS1 whose $z$-axis is along the line of sight
(see Figure~\ref{fig:precession}, top).  The precession axis is in the $y-z$ plane
inclined by angle $\psi$ to the plane of the sky\footnote{Positive values of
  $\psi$ indicate an inclination out of the plane of the sky towards the
  observer.}, and the jet axis makes an angle $\Theta$
with it.  For counter-clockwise\footnote{For the sense of rotation, we follow
  the convention that counter-clockwise rotation (as measured from the source
  along the precession axis) corresponds to a positive sign of the phase
  $\phi$.} precession, the coordinates of material at distance $r$ 
from the origin are 
\begin{equation}
  \left( \begin{array}{l} x \\ y \\ z \end{array} \right) =
       r \times \left(
                \begin{array}{l}
                   \sin\Theta\cos\phi \\
                   \cos\Theta\cos\psi + \sin\Theta\sin\phi\sin\psi\\
                   -\cos\Theta\sin\psi + \sin\Theta\sin\phi\cos\psi\\
                 \end{array}
                 \right)
\end{equation}

where $\phi = \phi_0 + kr$ is the jet's azimuthal angle from the $x$-axis.
The initial phase $\phi_{0}$ can be taken as $0^\circ$ since the direction 
perpendicular to the methanol feature A seems to coincide with the eastern wall
of the outflow cavity. The PA $\alpha$ of the $y$-axis can be set from the
average angle of the fan-shaped region in the speckle images as
$\alpha=-30^\circ$.  Using ${\mathrm v} = 250~\mbox{km~s$^{-1}$}$, as measured 
by \citet{gau95} within the H66$\alpha$ recombination line, leaves as free
parameters $P_{\text{prec}}$, $\Theta$, $\psi$, and the sense of
rotation.
Trying to fit the orientation of the maser disk, the orientation of the
UC~\HII~region, and the position of the NIR blobs with these parameters
simultaneously, we find reasonable agreement with a precession period
$P_{\text{prec}}=280 \pm 10$~yrs, a precession angle $\Theta=40^\circ \pm
3^\circ$, a counter-clockwise sense of rotation, and small inclination
$\psi=0^\circ \pm 10^\circ$.  At larger inclination angles ($\gtrsim
10^\circ$) loops start to appear, significantly degrading the agreement.
In Figure~\ref{fig:precession}, we show the projected trajectory of the
proposed wandering jet with the thick line, whereas the thin lines give the
path obtained with a variation of $\pm 5^\circ$ in $\alpha$ and $\psi$,
resembling the finite width of the flow.

The analytic model presented in this section might suffice in order to get
a rough estimate of the precession parameters, although it does not take into
account the interaction of the flow with the ambient medium nor the excitation
and cooling of the ambient material. 

These parameters can be used to predict how the orientation
of the methanol maser disk changes with time.  Using the PA at the phase
$\phi_{0}=0^\circ$ as reference, one expects that the PA changes only
marginally (less than $1^\circ$) within 10~yrs.  A much more significant
change of $10^\circ$ ($20^\circ$) would be expected after 36~yrs
(50~yrs), which would be detectable with future VLBI observations.

\subsubsection{Numerical molecular hydrodynamic simulations}
\label{cha:jetsimulations}

\begin{figure*}[htbp]
  \centering
  \includegraphics[width=18cm]{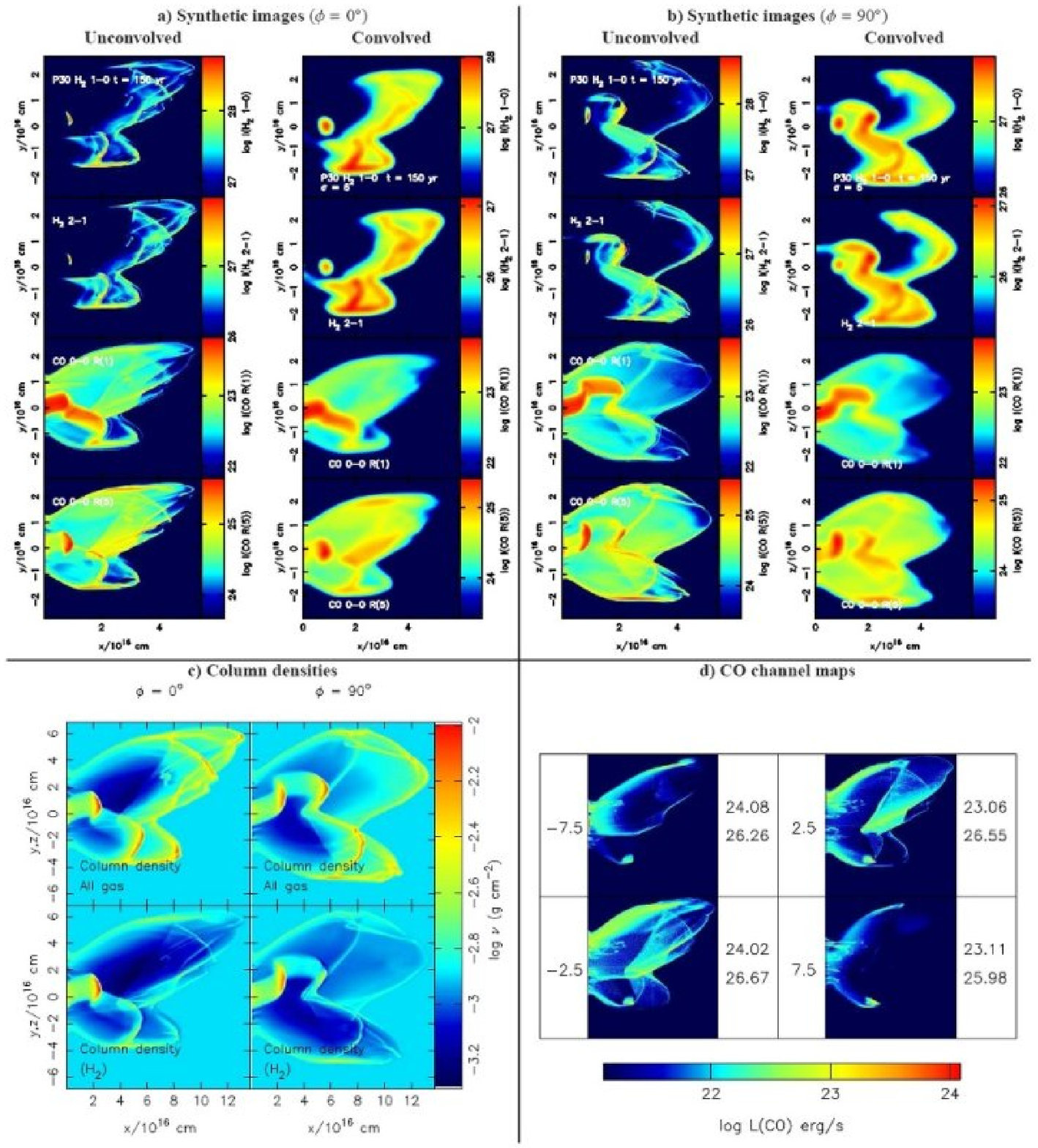}
  \caption{\textit{a)}, \textit{b)}: Synthetic images (H$_2$~$1\rightarrow0$, 
    $2\rightarrow1$; CO R(1), R(5)) from our ZEUS-3D molecular
     hydrodynamic simulation, shown for two different projections.
     In Figure~\textit{a)} the images are also shown after a convolution with a
     Gaussian which roughly resembles the resolution obtained in our speckle
     observation (Figure~\ref{fig:speckleimglabel}).
     Figure~\textit{c)} shows the total gas column density and the H$_2$
     column density for the same projections.
     Finally, in \textit{d)} we show channel maps of the CO outflow for four
     velocity bins. Each velocity bin has a width of 5~\mbox{km~s$^{-1}$} and
     the central velocity is given by the number on the left of each image (in
     \mbox{km~s$^{-1}$}). The two numbers on the right of each image indicate
     the log of the maximum integrated luminosity in any single element in the
     image (\textit{Top}, in erg~s$^{-1}$) and the log of the total integrated
     luminosity in the entire velocity bin (\textit{Bottom}).\newline
     The angle $\phi$ gives the angle between the $z$ axis and the LOS,
     corresponding to a rotation around the precession axis ($x$ axis).
     Please see the text (Sec.~\ref{cha:jetsimulations}) for
     a description of the complete model parameters. }
  \label{fig:mhd}
\end{figure*}

A large number of studies about the structure and evolution of precessing
protostellar jets can be found in literature~\citep[e.g.][]{rag93a, voe99,
  rag04, ros04, smi05a}, although most of these studies focus on jets from
low-mass stars with rather narrow precession angles.  As the number of
simulations carried out for wide precession angles is much more
limited~\citep[e.g.][]{cli96}, we performed a new hydrodynamic simulation.
Besides the general morphology, we aim for comparing the position of the
newly discovered fainter stars with the column density variations caused by a
precessing jet, which was beyond the scope of earlier studies. 

We use the version of the ZEUS-3D code as modified by~\citet{smi03},
which includes some molecular cooling and chemistry, as well as the ability
to follow the molecular (H$_2$) fraction.  
The large precession envisioned for the flow associated with NGC 7538 IRS1
requires that the simulation be performed on a very wide computational grid.
Due to computational limits, we were restrained to use for this simulation a
3D Cartesian grid of 275 zones in each direction, where each zone spans
$2\times 10^{14}$~cm in each direction.   This grid balances
the desire for some spatial resolution of the flow with the ability to
simulate a sufficiently large part of the observed flow associated with NGC
7538 IRS1. 
Still, the total grid size ($\sim 0.018$~pc) is smaller than the projected
distance between IRS 1 and \textit{K'}-band feature $C$.  

Owing to the rather small physical size of the grid, we have chosen a nominal
speed of 150~\mbox{km~s$^{-1}$}, reduced from the inferred value of
250~\mbox{km~s$^{-1}$} for this source.  The flow is precessed with a nearly
30$^{\circ}$ precession angle, with the amplitude of the radial components 
of the velocity 0.55 of that of the axial component.  The
precession has a period of 120~years, which leads to 1.25 cycles during a grid
crossing time. The flow is also pulsed, with a 30\% amplitude and a 30 year
period. This short period assists in the reproduction of the multiple knots of
\textit{K'}-band emission near NGC 7538 IRS1.  The jet flow also is sheared at
the inlet, with the velocity at the jet radius 0.7 that of the jet center.  We
have chosen a jet number density of 10$^5$ hydrogenic nuclei \mbox{cm$^{-3}$},
while the ambient density is 10$^4$.  The simulated jet radius is
$4.0\times10^{15}$~\mbox{cm} (20 zones). Thus, the time-averaged mass flux is
$2.6\times10^{-6}~M_{\sun}$~yr$^{-1}$, which is three orders of magnitude
lower than the value as determined for the CO outflow~\citep{dav98}. Similar
calculations of the momentum flux and kinetic energy flux, or mechanical
luminosity, yield values of
$3.8\times10^{-4}$~\mbox{km~s$^{-1}~M_{\sun}$~yr$^{-1}$} and 4.7~$L_{\sun}$, 
respectively. 

After the convolution with a PSF resembling the resolution obtained in real
observations the H$_2$~emission in our simulation shows a morphology which is
similar to the one seen in the \textit{K'}-band speckle image. In particular,
the simulations might also explain features $D$ and $E$ as associated with the
proposed precessing jet (compare Figure~\ref{fig:mhd}\textit{a} with
Figure~\ref{fig:speckleimglabel}).  The simulations also show that the CO emission,
which can be expected for a precessing jet at larger distances from the
driving source, appears very smooth, which is also in accordance with the
CO observations made for NGC~7538~IRS1.

We note that the fainter stars $e$ to $n$ reported in
Sec.~\ref{cha:lowmassstars} are located in the region where the column
density in our precessing jet simulation appears particularly low (see
Figure~\ref{fig:mhd}\textit{c}, left column), supporting the scenario
proposed in Sec.~\ref{cha:precindications}.

\subsubsection{Possible precession mechanisms}

Several mechanisms have been proposed which can cause jet bending or jet
precession, although most of them were established for low- and intermediate
stars and can cause precession angles of only a few degrees~\citep{fen98,
  eis97}.  For the case of high-mass stars and larger precession angles
($\Theta \sim 40^\circ$), \citet{she00} summarized the three most promising
concepts that could induce precession into circumstellar disks.  
We discuss how well these mechanisms can explain the observations of IRS1.
For all cases, it is assumed that the outflow is launched close to the center
of the disk and that a precession of the inner parts of the disk will
translate into a precession of the collimated flow~\citep{bat00}.\\[3mm]
\noindent
\textbf{1.~Radiative-induced warping:}
  \citet{arm97} suggested that geometrically
  thin, optically thick accretion disks can become unstable to warping if the
  incident radiation from the stellar source is strong enough.  As this
  warping instability is expected to occur only at disk radii larger than a
  critical radius $R_{\text{crit}}$, we can estimate whether
  radiative-induced disk warping is expected at the inner part of the IRS1
  disk.  Using a stellar mass of $M \sim 30~M_{\sun}$~\citep{pes04}, a mass
  accretion rate of the order of the mass outflow rate $\dot{M}_{\text{acc}}
  \approx \dot{M}_{\text{outflow}} \approx 5.4 \times
  10^{-3}~M_{\sun}~\text{yr}^{-1}$~\citep{dav98}, and a luminosity $L \approx
  9.6 \times 10^4~L_{\sun}$~\citep{aka05}, we use equation 5 by~\citet{arm97}
  and the assumptions listed in~\citet{she00} and obtain a critical radius
  $R_{\text{crit}} \sim 200$~pc.
  Since this is far beyond the inner edge of the disk where the jet
  collimation is expected to happen, it is very unlikely that the radiation
  emitted by the star or due to accretion causes any noticeable warping within
  the disk.\\[3mm]
\noindent
\textbf{2.~Anisotropic accretion events:}
  The impact/merging of (low mass) condensations can change the orientation of
  the disk angular momentum vector.  In such a dramatic event, angular
  momentum can be transferred from the impactor onto the accretion disk,
  potentially resulting in a net torque in the rotation of the disk.  To
  estimate the precession angle, which could result from anisotropic
  accretion, very detailed assumptions about the disk, the impacting
  condensation, and their kinematics must be made.  Since no data is available
  to estimate these quantities, we refer to the example computed
  by~\citet{she00} and note that in extreme cases, such an accretion event
  could cause a sufficiently large precession angle in the case of 
  NGC~7538~IRS1 as well.  However, in this scenario one would expect rather
  sudden changes in the jet direction rather than a smooth precession.\\[3mm]
\noindent
\textbf{3.~Tidal interactions with a companion:}
  Warping and precession of
  the disk could be caused by tidal interactions with one or more companions
  on non-coplanar orbits.  We assume the simplest case of a binary: with
  stellar masses $M_p$ (primary) and $M_s$ (secondary), an orbit with
  inclination $i$ with respect to the disk plane, and a semimajor axis $a$.
  The mass ratio shall be denoted $q=M_p/M_s$ and will be assumed as unity.
  Our observations place an upper limit on the separation of such a companion
  (see Figure~\ref{fig:IRS1companionlimits}).  Two cases can be
  considered:\newline
  \textbf{\boldmath 3a) circumprimary disk ($a>r_o$): \boldmath}~Because tidal torques would
  truncate the disk at about 0.3 times the binary separation~\citep{lub00}, we
  obtain a lower limit for the binary separation (for a circular orbit),
  namely, $a>2\,500$~AU.
  However, a binary with such a large separation would be not suited to
  explain the observations since the orbital period would be $> 2 \times
  10^4$~yrs ($M_p+M_s=30~M_{\sun}$), implying a disk precession rate of
  $> 4 \times 10^5$~yrs~\citep{bat00}.  Assuming an extreme eccentricity
  might yield a short precession period of the order of $10^2$~yrs
  but implies strong, periodic interactions between the companion
  and the disk during each perihelion passage.  As this would quickly
  distort and truncate the disk, we see this assumption contradicts
  the methanol maser structure, which suggests a smooth extension of the
  methanol layer from $\sim 290$~AU to $\sim 750$~AU.\newline
  \textbf{\boldmath 3b) circumbinary disk ($a<r_i$): \boldmath}~Smoothed particle
  hydrodynamic simulations by~\citet{lar97} showed that a binary on a
  non-coplanar orbit with large inclination $i$ could cause strong
  quasi-rigid body precession of the circumbinary disk (for $q > 10$) and
  strong warping, especially on the inner edge of the disk ($q \sim 1$).
  The same authors report that the disk precession frequency
  $\omega_{\text{prec}}$ should be lower than the orbital frequency of the
  binary $\omega_{\text{binary}}$.  To make an order-of-magnitude estimation
  for the orbital period that would be expected for this hypothetical IRS1
  binary system, one can assume $\omega_{\text{prec}} \approx
  \omega_{\text{binary}} / 20$~\citep{bat00} to obtain $P_{\text{binary}}\approx
  14$~yrs for the binary period, corresponding to a separation of
  $a_{\text{binary}} \approx 19$~AU ($\sim 7$~mas).  This binary separation
  then puts a lower limit on the radius of the inner edge of the circumstellar
  disk.  As this scenario can trigger the fast disk precession without
  truncating the extended disk structure traced by the methanol masers, we
  consider a circumbinary disk as the most plausible explanation.\\

\subsection{The IRS2 companion and flow interaction with the IRS2~UC~\HII~region}

The spectral type of IRS2 was estimated to be O4.5~\citep{aka05}, corresponding
to a luminosity of $\sim 6.4 \times 10^{5} L_{\sun}$.  Using the measured $K'$-band
flux ratio, one can make rough estimates for the spectral type of the two
components reported in Sec.~\ref{cha:IRS2binarity}.  By assuming the total
luminosity is attributed only to the two components, we obtain a spectral type of
O5 for IRS2$a$ and O9 for IRS2$b$ (using the OB star luminosities
from~\citealt{vac96}).\\

Within our images, the wide-opening angle outflow cone from IRS1 seems to
extend well out to IRS2.  This offers an explanation for the shock tracer
line emitting region that was imaged around IRS2~\citep[][ see
Figure~\ref{fig:mosaic}$i$]{blo98}.  The bowshock-like morphology of the
\FFeII~and H$_2$ emission suggests that the shock is excited from the south
(which is roughly the direction towards IRS1). In the direction opposite IRS1,
the \FFeII~and H$_2$ emission even shows a cavity-like structure, which also
appears in the 6~cm-radio continuum. \citet{blo98} suggested a stellar wind
bowshock scenario, in which IRS2 moves with a speed of
$\sim$10~\mbox{km~s$^{-1}$} towards the southwest through the ambient
molecular cloud.  We note that the morphology could also be explained by
interaction between the IRS1 outflow and IRS2 outflows. Based on its
young age, IRS2 might also launch a powerful wind itself, causing the distinct
shock zone which appears within the shock tracer emission (see
Figure~\ref{fig:mosaic}\textit{i}) and which is also detectable in our
\textit{K'}-band image (arc-like morphology between features $G$ and $H$).

\subsection{Outflow structures from IRS1 at larger spatial scales}

Figure~\ref{fig:mosaic}\textit{a} to \textit{d} shows mosaics of the vicinity
of IRS1 in the four IRAC bands.  Although IRS1 and IRS2 appear saturated in
these images (shown in logarithmic scaling) and banding (vertical and
horizontal stripes produced by IRS1 and IRS9) appears especially in the 5.8
and $8.0~\mu$m images, structures potentially related to IRS1 can be
observed. About 40\arcsec~towards the southeast of IRS1, a bowshock structure
can be seen, which is also present in the H$_2$ image by \citet{dav98}.  This
bowshock points in the same direction as the redshifted lobe of the CO
outflow (see Figure~\ref{fig:mosaic}\textit{f}) and just opposite to the
outflow direction identified in our speckle image at small scales.  Thus,
it is possible that this bow traces the southeastern part of the IRS1 outflow.
In our speckle image, the inner part of this southeastern outflow is not
visible; likely a result of strong intervening extinction.

Furthermore, it is interesting to note that the ``ridge'' connecting the
IRS1--3 cluster with IRS4 and IRS5 follows the western wall of the outflow
direction identified in our speckle image (see top of
Figure~\ref{fig:mosaic}).  It is possible that the total extent of the IRS1
outflow also reaches much further northwest than the structure
seen in the speckle image, contributing to the excitation of the western part
of the bubble seen in the IRAC bands and the shocks in the H$_2$
image by \citet[][ see Figure~\ref{fig:mosaic}$h$]{dav98}.

\section{Evidence for triggered star formation in the NGC~7538 star forming
  region}

It has been proposed by many authors that star formation seems
to propagate southeastwards throughout the NGC~7538 complex (\citealt{wer79,
mcc91, ojh04}). This is indicated by the spatial arrangement of the
members of this star forming region, which also seems to agree with the
expected evolutionary sequence: starting about $3'$ northwest of IRS1,
O~stars located in the \HII~region represent the most developed evolutionary
state, followed by the IRS1--3 cluster and their associated UC\HII~regions,
with the compact reflection nebula around IRS9 representing the youngest
member of this star formation site.  In agreement with this picture,
\citet{bal04} measured the reddening of stars throughout NGC~7538 and found a
gradient in reddening with the most heavily reddened sources in the
southeast.\\

The presented IRAC images can also be interpreted in support of this scenario
since the "ridge"-like feature connecting IRS1-3, IRS4, IRS5 seems to trace the
interface between the northeastern bubble (visible at NIR/MIR wavelengths)
and the submillimeter bubble, which appears in the 450~$\mu$m and
850~$\mu$m-maps by~\citet[][ see Figure~\ref{fig:iracirs}]{rei05}. This
suggests that in NGC~7538, star formation was triggered by the compression
of gas just at the interface layer of these expanding bubbles,
sequentially initiating the formation of the observed chain of infrared sources.\\

\citet{ojh04} suggested that IRS6, the most luminous source in the NGC~7538
region, might be the main exciting source responsible for the optical
\HII~region.  Inspecting the IRAC color composites, this scenario is supported
by the morphology of the bright, curved structure west of IRS6.
In the IRAC 8~$\mu$m band (red in Figure~\ref{fig:iracirs}$c$), this structure
appears particularly prominent.  As it is known that emission in this IRAC band
is often associated with PAHs, this suggests that this region is illuminated by
strong UV radiation from IRS6. Other features, such as the conical structure
around 2MASS\,23135808+6130484 and the structure northeast of IRS7, also show
a symmetry towards IRS6.

%

\section{Summary and Conclusions}

Bispectrum speckle interferometry and archival \textit{Spitzer}/IRAC
imaging of the massive protostars NGC~7538 IRS1/2 and their vicinity are
presented.  We summarize our results as follows:

\begin{enumerate}
\item The clumpy, fan-shaped structure seen in our speckle images most likely traces
  recent outflow activity from IRS1, consistent with the direction
  of the blue-shifted lobe of the known CO outflow.  A bowshock structure
  noticeable in the IRAC images $\sim 40$\arcsec~southeast of IRS1 suggests
  that the total extent of the outflow might be several parsecs.  The outflow
  might have also contributed to shaping and exciting the bubble-like
  structure, which is prominent in all four IRAC bands (although
  contributions from several other sources, especially IRS6, are also evident).

\item A companion around the high-mass star NGC~7538 IRS2 was discovered.
  Furthermore, we see indications for interactions between the IRS1 flow
  and outflows or stellar winds from IRS2 (nebulosity surrounding IRS2).

\item A jet precession model seems suitable to describe the features
  observed within our NIR images, simultaneously explaining the
  misalignment between the putative methanol maser disk, the UC\HII~region,
  and the outflow tracers detected at larger scales (CO).  A simple analytic
  precession model was used to extract order-of-magnitude estimates for the
  precession parameters.  Using these we estimate tidal interaction of a
  close binary system with a circumbinary disk as the most plausible
  gyroscopic mechanism, which is triggering the precession.
  
\item  The presented molecular hydrodynamic simulations can reproduce some of
  the fine-structure observed in our NIR images and indicate that the
  arrangement of the detected fainter stars might be explained as a
  column-density effect, caused by the proposed precessing jet.

\item The prominent sites of ongoing high-mass star formation in NGC~7538 seem to
  be located just at the interface between two bubble-like structures --- one
  is visible in the presented IRAC images, the other traced by submillimeter
  observations.   The gas compression caused by the expansion of these bubbles
  might have triggered star formation in this region.
\end{enumerate}

While it is well established that the outflows of HMPOs generally appear less
collimated than those of their low-mass counterparts, the recent discovery of
evidence for outflow precession for an increasing number of massive YSOs might
indicate a common launching mechanism for all outflow driving sources of
all stellar masses.  The observed widening in HMPO outflows might be due to
selection effects~\citep{she05} and/or precession of a collimated jet.  The
large precession angles reported for IRAS~20126+4104~\citep{she00},
S140~IRS1~\citep{wei02}, IRAS 23151+5912~\citep{wei06}, and now NGC~7538~IRS1
(this paper) might point towards a rather dramatic precession mechanism, maybe
the presence of very close, high-mass companions on non-coplanar orbits.

We strongly encourage further observations of IRS1, especially to
detect potential companions either by near-infrared long-baseline
interferometry or radial velocity measurements.

\begin{acknowledgements}
We thank all the participants of the \textit{NGC~7538 collaboration} for very
fruitful discussions that contributed to the achievement of this paper.  The
collaboration consists of Roy Booth, John Conway, James De Buizer, Moshe
Elitzur, Stefan Kraus, Vincent Minier, Michele Pestalozzi, and Gerd Weigelt.\newline
We also acknowledge the BTA and MMT staff for their support of this run,
and D.~Apai and I.~Pascucci for assistance during the MMT observations.\newline
SK was supported for this research through a fellowship from the International
Max Planck Research School (IMPRS) for Radio and Infrared Astronomy at the
University of Bonn.\newline
The numerical hydrodynamic simulations were executed on the Armagh SGI Origin
2000 computer (FORGE), acquired through the Particle Physics and Astronomy
Research Council (PPARC) JREI initiative with SGI participation.\newline
This work is based in part on archival data obtained with the Spitzer Space
Telescope, which is operated by the Jet Propulsion Laboratory, California
Institute of Technology under a contract with NASA.\newline
This publication makes use of data products from the Two Micron All Sky
Survey, which is a joint project of the University of Massachusetts and the
Infrared Processing and Analysis Center/California Institute of Technology,
funded by the National Aeronautics and Space Administration and the National
Science Foundation.
\end{acknowledgements}

\bibliographystyle{aa}
\bibliography{5068}

\end{document}